\documentclass[useAMS,usenatbib,referee]{biom}
\def\bSig\mathbf{\Sigma}

\usepackage[figuresright]{rotating}
\usepackage{amsmath}
\usepackage{multirow}
\usepackage{multicol}
\usepackage{booktabs}
\usepackage{ifthen}
\usepackage{xcolor}

\title[Sparse two-stage Bayesian meta-analysis for individualized treatments]{Sparse two-stage Bayesian meta-analysis for individualized treatments}

\author{Junwei Shen$^*$\email{junwei.shen@mail.mcgill.ca}, 
Erica E.M. Moodie, and Shirin Golchi \\
Department of Epidemiology, Biostatistics and Occupational Health, \\McGill University, Montréal, Québec, Canada}

\begin{document}


\date{{\it Received October} 2007. {\it Revised February} 2008.  {\it
Accepted March} 2008.}



\volume{64}
\pubyear{2008}
\artmonth{December}


\doi{10.1111/j.1541-0420.2005.00454.x}




\begin{abstract}
Individualized treatment rules tailor treatments to patients based on clinical, demographic, and other  characteristics. Estimation of individualized treatment rules requires the identification of individuals who benefit most from the particular treatments and thus the detection of variability in treatment effects. To develop an effective individualized treatment rule, data from multisite studies  may be required due to the low power provided by smaller datasets for detecting the often small treatment-covariate interactions. However, sharing of individual-level data is sometimes constrained. Furthermore, sparsity may arise in two  senses: different data sites may recruit from different populations, making it infeasible to estimate identical models or all parameters of interest at all sites, and the number of non-zero parameters in the model for the treatment rule may be small. To address these issues, we adopt a two-stage Bayesian meta-analysis approach to estimate individualized treatment rules which optimize expected patient outcomes using multisite data without disclosing individual-level data beyond the sites. Simulation results demonstrate that our approach can provide consistent estimates of the parameters which fully characterize the optimal individualized treatment rule. We estimate the optimal Warfarin dose strategy using data from the International Warfarin Pharmacogenetics Consortium, where data sparsity and small treatment-covariate interaction effects pose additional statistical challenges. 
\end{abstract}

%

\begin{keywords}
Bayesian meta-analysis; Individualized treatment rules; Multisite studies; Personalized medicine.
\end{keywords}


\maketitle

\section{Introduction}
\label{sec1}
In the traditional one-size-fits-all approach, patients with the same disease receive the same treatment regardless of their individual characteristics.  This strategy can be suboptimal, as the best treatment for one patient might be ineffective for another due to treatment effect heterogeneity among patient subgroups.  The personalized medicine approach has emerged as an alternative, leveraging this treatment effect heterogeneity in clinical decision-making to recommend the most appropriate treatment to individual patients~\citep{chakraborty2013statistical}. Dynamic treatment regimes (DTRs), which consist of a sequence of decision rules, formalize the statistical framework of personalized medicine in the setting of multiple treatment stages~\citep{chakraborty2013statistical,chakraborty2014dynamic,laber2014dynamic}. When only a single treatment stage is considered, a DTR reduces to an individualized treatment rule (ITR), which is the focus of our work.


Several methods have been proposed to estimate optimal ITRs (i.e., rules to optimize expected patient outcomes) or their multi-stage counterparts for various outcome or exposure types based on individual-level data. 
Among the many alternatives, regression-based approaches indirectly estimate the optimal ITR by first modelling the expected outcome as a function of treatment, covariates, and their interactions and then determining the treatment that, for each covariate combination, will optimize the estimated expected outcome. Common regression-based approaches include Q-learning~\citep{watkins1989learning,sutton2018reinforcement}, G-estimation~\citep{robins2004optimal}, and dynamic weighted ordinary least squares (dWOLS)~\citep{wallace2015doubly}. One practical challenge of ITR estimation is the low power for detecting treatment-covariate interactions~\citep{greenland1983tests}. To increase the sample size and generalizability of the findings, multisite data are attractive. Ideally, in a multisite study one could pool the individual-level data from different sites together and analyze the pooled data in a central analysis site. However, this may be infeasible due to, e.g., institutional policies which restrict the sharing of individual-level information. 
Therefore, it is desirable to develop valid methods for ITR estimation that avoid sharing individual records. 

Different strategies have been proposed for analyses under constrained data sharing~\citep{rassen2013evaluating}, but few have focused specifically on restrictions on data sharing in the context of ITR estimation~\citep{danieli2022preserving,moodie2022privacy} and none account for
the possibility that parameters of interest may vary across centers or sites.
Meta-analysis, a widely-used approach for evidence synthesis
, can avoid releasing individual-level records when analyzing multisite data~\citep{rassen2013evaluating}. However, classic meta-analysis techniques based on aggregate data may be unsuitable for identifying treatment effects within subgroups under heterogeneity~\citep{berlin2002individual, simmonds2007covariate}, which is the target of personalized medicine, and so it is unclear whether a meta-analysis approach to ITR estimation is feasible. While using individual participant data (IPD) for meta-analysis~\citep{doi:https://doi.org/10.1002/9781119333784.ch7} appears promising for ITR analysis, a one-stage IPD meta-analysis requires combining all individual-level data into a single dataset, and thus cannot be used in settings where individual records cannot be shared and thus a two-stage approach is required.  In this work, we adopt a Bayesian two-stage IPD meta-analysis approach to estimate the optimal ITR using multisite data without the need for sharing individual-level information across sites. We note that treatment-covariate interactions sometimes are described as ``treatment effect heterogeneity'' in the literature. However, in the meta-analysis literature, heterogeneity typically refers to the variability across sites. To avoid confusion, from now on, we adopt the meta-analysis tradition and reserve the word \textit{heterogeneity} for variability across sites. 

Warfarin is a widely-used oral anticoagulant for thrombosis and thromboembolism treatment and prevention \citep{rettie2006pharmocogenomics}. Establishing an optimal Warfarin dose strategy is of vital importance due to the narrow therapeutic window and the large interindividual variability in patients' response to the drug \citep{rettie2006pharmocogenomics,international2009estimation}. The data from the International Warfarin  Pharmacogenetics Consortium was collected in 9 countries from 4 continents. These data have suggested 
factors that are possibly associated with Warfarin dosing~\citep{international2009estimation}. However, statistical challenges arise in the context of a two-stage Bayesian meta-analysis of the optimal Warfarin dose strategy. 

One challenge is data sparsity, a term we use to refer to the phenomenon of not observing a sufficient number of patients with a given set of characteristics. For example, race is potentially a tailoring variable for Warfarin dosage. In certain sites all patients fall under only one category of race (e.g., White) and therefore inference about the interaction between race and Warfarin dose cannot be made based on the data from these sites. 
A na\"ive approach would remove sites with sparse data, leading to a significant loss of information. 

A second challenge is model sparsity, i.e., extremely small (practically zero) treatment-covariate interactions resulting from variables that are irrelevant for the dosing decision. The small effect estimates can lead to invalid dose recommendations (e.g., outside of the appropriate range) or the estimated optimal dose for an individual being highly volatile or sensitive to the parameter estimates. Including covariates with no tailoring effect also makes the estimated optimal dosing strategy unnecessarily complex.

We address data sparsity by reparametrizing the likelihood such that the site-specific estimates are linked to the correct set of parameters. To address the second challenge, 
we use shrinkage priors~\citep{van2019shrinkage}, opting for a horseshoe prior due to its proven advantages in maintaining large effects and efficiently  handling sparsity~\citep{shrinkage}. 
Methods including the notation and assumptions are described in section \ref{sec2}. In section \ref{sec3}, we explore the performance of the proposed method across a range of scenarios via simulation studies. In section \ref{sec4}, we estimate an optimal individualized Warfarin dose strategy without the need for sharing patient-level data. The paper concludes with a discussion.  

\section{Methods}\label{sec2}
\subsection{Preliminaries}
\label{subsec21}
Let $Y$ denote a continuous outcome of interest, where larger values of $Y$ are preferable. Let $A$ denote the binary or continuous treatment received by the patient, and  $\textbf{\textit{X}}$ be a vector of  pre-treatment covariates. Let $Y^a$ be the potential outcome a patient would experience if assigned treatment $a$. Uppercase, lowercase, and bold denote random variables, realizations of random variables, and vectors respectively. 

An ITR $d(\textbf{\textit{X}}): \textbf{\textit{X}} \rightarrow A$ tailors treatment to patients based on individual characteristics, $\boldsymbol{X}$. An optimal ITR $d^{opt}(\textbf{\textit{X}})$ maximizes the value function under $d^{opt}(\textbf{\textit{X}})$, that is, the expected potential outcome $E(Y^{d(\textbf{\textit{X}})})$ if all patients in a population are treated according to $d(\textbf{\textit{X}})$. Identification of an optimal ITR 
relies on several assumptions: (i) the stable unit treatment value assumption (SUTVA): a patient’s outcome is not influenced by other patients’ treatment~\citep{rubin1980discussion}; (ii) no unmeasured confounding \citep{robins1997causal}; (iii) positivity: $p(A=a\vert \boldsymbol{X} = \boldsymbol{x}) > 0$ almost surely for all possible $\boldsymbol{x}$ and $a$~\citep{10.1093/aje/kwn164}. Additional modelling assumptions will also be required for the (regression-based) meta-analytic approach that we pursue. 

To identify the optimal ITR,  an outcome model could be specified and decomposed into two components: 
 $   E(Y\vert A=a,\textbf{\textit{X}}=\textbf{\textit{x}}) = f(\textbf{\textit{x}}^{\boldsymbol{(\beta)}};\boldsymbol{\beta}) + \gamma(a, \textbf{\textit{x}}^{\boldsymbol{(\psi)}}; \boldsymbol{\psi})$,
where both $\textbf{\textit{x}}^{\boldsymbol{(\beta)}}$ and $\textbf{\textit{x}}^{\boldsymbol{(\psi)}}$ are subvectors of $\boldsymbol{x}$, and include predictive covariates and covariates that interact with treatment (prescriptive variables), respectively. 
The prescriptive covariate vector $\textbf{\textit{x}}^{\boldsymbol{(\psi)}}$ is a subvector of $\textbf{\textit{x}}^{\boldsymbol{(\beta)}}$. 
The parameter vectors in the treatment-free function $f(\textbf{\textit{x}}^{\boldsymbol{(\beta)}};\boldsymbol{\beta})$ and the blip function $ \gamma(a, \textbf{\textit{x}}^{\boldsymbol{(\psi)}}; \boldsymbol{\psi})$~\citep{robins2004optimal} are denoted by $\boldsymbol{\beta}$ and $\boldsymbol{\psi}$, respectively.  The treatment-free function
depends on covariates $\textbf{\textit{x}}^{\boldsymbol{(\beta)}}$ but not treatment, and thus is not relevant to treatment or dosing decisions. 
The difference in the expected potential outcome of patients receiving treatment $A=a$ and reference $A=0$, with the same prescriptive covariates $\textbf{\textit{X}}=\textbf{\textit{x}}^{\boldsymbol{(\psi)}}$,  is represented by the blip function:
 $   \gamma(a,\textbf{\textit{x}}^{\boldsymbol{(\psi)}}; \boldsymbol{\psi}) = E(Y^a -Y^0\vert \textbf{\textit{X}}=\textbf{\textit{x}}^{\boldsymbol{(\psi)}}; \boldsymbol{\psi})$. 
The blip function satisfies $\gamma(0, \textbf{\textit{x}}^{\boldsymbol{(\psi)}}; \boldsymbol{\psi})=0$ and the optimal ITR is then defined as $d^{opt}(\textbf{\textit{x}}) = \text{arg}\max_a \gamma(a, \textbf{\textit{x}}^{\boldsymbol{(\psi)}};\boldsymbol{\psi})$, since the treatment assignment influences the expected outcome only through the blip function. Therefore, estimation of the optimal ITR requires correct specification of the blip function and estimation of the blip parameter $\boldsymbol{\psi}$. 
A common form for the blip function in the binary treatment setting $A \in \{0,1\}$ is $\gamma(a, \textbf{\textit{x}}^{\boldsymbol{(\psi)}};\boldsymbol{\psi}) = a\textit{g}(\boldsymbol{x}^{\boldsymbol{(\psi)}};\boldsymbol{\psi})$, and linear models could be assumed for both \textit{f} and \textit{g}. Under a linearity assumption, the outcome model becomes 
\begin{align}
    E(Y \vert A=a, \boldsymbol{X} = \boldsymbol{x}) = \boldsymbol{\beta^Tx^{(\beta)}} + a\boldsymbol{\psi^Tx^{(\psi)}},
    \label{eq1}
\end{align}
where, in this form, we assume that $\boldsymbol{x^{(\beta)}}$ and $\boldsymbol{x^{(\psi)}}$ have been augmented by a column of 1s to ensure an intercept and main effect of treatment, respectively. 
In this setting, the optimal ITR is given by $d^{opt}(\boldsymbol{x}) = I(\boldsymbol{\psi^T x^{(\psi)}}>0)$. For a continuous treatment (e.g., a dose of a drug), a blip function should be specified so that the optimal treatment can be an interior point of the set of possible treatments.  For example, a quadratic or higher order term for the treatment could be included in the blip function: 
$\gamma(a,\textbf{\textit{x}}^{\boldsymbol{(\psi)}};\boldsymbol{\psi}) = (\boldsymbol{\psi^{(1)T}}, \boldsymbol{\psi^{(2)T}}) (a\textbf{\textit{x}}^{\boldsymbol{(\psi^{(1)}})},  a^2\textbf{\textit{x}}^{\boldsymbol{(\psi^{(2)}})})^T$.

Various approaches are available for unbiased and consistent estimation of the blip parameter $\boldsymbol{\psi}$. For example, basic Q-learning can fit a standard linear regression to the model in equation (\ref{eq1}) or indeed a more flexible model; consistency of the estimation is guaranteed under correct model specification \citep{chakraborty2013statistical}. 
Other regression-based methods such as G-estimation and dWOLS, both of which are doubly robust, could also be employed~\citep{robins2004optimal,wallace2015doubly}. Bayesian approaches have also been proposed, such as Bayesian G-computation~\citep{ArjasSaarela+2010}, a Bayesian machine learning approach to Q-learning~\citep{murray2018bayesian}, Bayesian additive regression trees~\citep{logan2019decision}, and Bayesian causal forest~\citep{hahn2020bayesian}.

\subsection{Two-stage IPD meta-analysis}\label{subsec22}

We now describe a two-stage IPD meta-analysis approach in the Bayesian framework to estimate the optimal ITR when multisite data are available but data sharing across sites is not allowed. Let $K$ be the number of sites. For illustration purposes, we assume linear models for both treatment-free and blip functions. The extension to other parametric outcome models is straightforward. The site-specific outcome model can be written as: 
 $   E(Y_{ij} \vert \boldsymbol{X} =\boldsymbol{x_{ij}}, A = a_{ij}) = \boldsymbol{\beta_i^Tx_{ij}^{(\beta)}} + a_{ij} \boldsymbol{\psi_i^T x_{ij}^{(\psi)}}$,
where $i \in \{1, \ldots, K\}$ and $j \in \{1, \ldots, n_i\}$ index the site and individual patient in a given site respectively, and $n_i$ is the number of patients in site $i$.
The predictive and prescriptive covariate vectors are denoted by $\boldsymbol{x_{ij}^{(\beta)}}$ and $\boldsymbol{x_{ij}^{(\psi)}}$, respectively. The $p$-dimensional site-specific treatment-free parameter $\boldsymbol{\beta_i}=(\beta_{i0},\ldots,\beta_{i,p-1})$ and $q$-dimensional blip parameter $\boldsymbol{\psi_i}=(\psi_{i0},\ldots, \psi_{i,q-1})$ have similar interpretations to those in equation (\ref{eq1}), except that the target here is the site-specific ITR.  The site-specific parameters $\boldsymbol{\beta_i}$ and $\boldsymbol{\psi_i}$ may not be identical across sites. We initially assume that the specific variables included in the vectors $\boldsymbol{x^{(\beta)}_{ij}}$, $\boldsymbol{x_{ij}^{(\psi)}}$ are identical across sites; we later relax that assumption. 

Suppose that our interest is not in the site-specific optimal ITRs but a common optimal ITR that could be applied to all sites and, more generally, to future patients at comparable sites that may not have contributed data to the estimation. When the site-specific parameters are not identical across sites, it may be reasonable to assume that a common distribution exists for the varying site-specific blip parameters:
\begin{align}
    \boldsymbol{\psi_i} \sim MVN(\boldsymbol{\psi}, \boldsymbol{\Sigma_\psi}),
    \label{eq2}
\end{align}
where MVN represents the multivariate normal distribution, and $\boldsymbol{\psi}=(\psi_0,\ldots,\psi_{q-1})$ and $\boldsymbol{\Sigma_\psi}$ are the common mean vector and variance-covariance matrix, respectively.

The two-stage IPD meta-analysis approach estimates a common optimal ITR by first conducting separate analyses of site-specific optimal ITRs and then combining the site-specific optimal ITRs via a hierarchical model. Specifically, at the first stage, each site obtains estimates of blip parameters $\hat{\psi}_{it}$ and the associated standard deviations sd$(\hat{\psi}_{it})$, for $t = 0, \ldots, q-1,$  
As mentioned in the last section, $\hat{\psi}_{it}$ and sd$(\hat{\psi}_{it})$ can be obtained from various approaches such as Q-learning or dWOLS, using only site-specific data. At the second stage, only those site-specific estimates will be transferred to a central analysis site (and thus the individual-level data are preserved), and combined in a Bayesian hierarchical model: 
\begin{equation}
\begin{split}
     \hat{\psi}_{it} &\sim N(\psi_{it}, \text{sd}(\hat{\psi}_{it})^2),  \quad 
     \psi_{it} \sim N(\psi_t, \sigma_{\psi_t}^2),\\
     \psi_t & \sim p_{\psi_t}(\psi_t), \quad 
     \sigma_{\psi_t}^2 \sim p_{\sigma_{\psi_t}^2}(\sigma_{\psi_t}^2).
     \end{split}  
     \label{eq:twostage}
\end{equation}
Here, $\psi_{it}$ and $\psi_t$ are the $(t+1)$-th elements of the site-specific and common blip parameter vectors. The between-site heterogeneity associated with $\psi_{it}$ is denoted by $\sigma_{\psi_t}^2$. 
Prior distributions  $p_{\psi_t}$ and $p_{\sigma_{\psi_t}^2}$ can be assigned for the unknown parameters $\psi_t$ and $\sigma_{\psi_t}^2$. Popular prior choices include a normal prior with large variance for the mean parameter $\psi_{t}$ and a half-Cauchy prior for the variance component parameter $\sigma_{\psi_t}$~\citep{gelman2006prior,gelman2013bayesian}. We use these prior choices in our simulation studies and  the optimal Warfarin dosing analysis that follows. The Bayesian hierarchical model can be easily fitted in any Bayesian software. In this paper, we use \texttt{RStan} \citep{rstan,stan}. 

The two-stage IPD meta-analysis approach requires each site to provide estimates of blip parameters and the associated standard deviations only, which avoid sharing the individual-level data. At the first stage, the treatment-free parameters are estimated together with the blip parameters. However, they are irrelevant for the optimal treatment decision and will not be used in the second stage. 
We make no common distribution assumptions of site-specific treatment-free parameters in (\ref{eq2}). In reality, if the blip parameters come from a common distribution, then it is highly likely that the treatment-free parameters are also from a common distribution. However, this is not required for our approach. The model at the second stage depends on unbiasedness, consistency, and normality of site-specific estimates from the first stage. 
Therefore, alternative stage-one models could be considered for different site-specific ITRs, as long as unbiasedness, consistency, and normality are assured. In the following sections, we assume no model misspecification exists. 
We establish the link between the proposed two-stage approach and a one-stage approach based on full individual-level data in Web Appendix S1 of the Supplementary Materials.

\subsection{Sparsity}
\label{subsec24}
So far, we have assumed that the specific variables included in the vectors $\boldsymbol{x_{ij}^{(\beta)}}$, $\boldsymbol{x_{ij}^{(\psi)}}$ are identical across sites. However, 
estimation of identical models at all sites may be infeasible due to heterogeneity of patient populations across sites (data sparsity). In addition, when a large number of covariates are available, the number of non-zero parameters in the model may be small (model sparsity). We now describe our approach to sparsity. 

\subsubsection{Data sparsity}
Data sparsity occurs when insufficiently many patients with a given set of characteristics are represented in the samples at all sites. Then, not all site-specific parameters can be estimated for sites with sparse data. We employ a Bayesian hierarchical model which borrows information across sites; this requires modification of likelihood contribution for sites with data sparsity.   

To illustrate this, consider a toy example, with a binary covariate $X\in\{0,1\}$ (i.e., $p=q=2$), and the following true outcome model for an individual at site $i$: 
  $  E(Y\vert X) = \beta_{i0} + \beta_{i1} X + A(\psi_{i0} + \psi_{i1} X)$.
Therefore, $\psi_{i0}$ is the difference in the mean outcome between $A=1$ and $A=0$ when $X=0$ 
in site $i$; $\psi_{i1}$ is the difference in the treatment effect between patients with $X=1$ and $X=0$ in site $i$; $\psi_{i0}+\psi_{i1}$ is the treatment effect for patients with $X=1$ in site $i$. Then we consider two scenarios for data sparsity.

In the first scenario, $X=0$ for all patients in site $i$. In this case, the main effect of $X$, $\beta_{i1}$, and the treatment-covariate interaction $\psi_{i1}$ cannot be estimated. The site-specific outcome model reduces to 
          $E(Y\vert X) = \gamma_{i0} + A\xi_{i0}$. 
    Since this model 
    is fitted among patients with $X=0$, $\xi_{i0}$ is the treatment effect for patients with $X=0$. That is, $\xi_{i0}=\psi_{i0}$. 
    The likelihood contribution of site $i$ is then 
    $\hat{\xi}_{i0} \sim N(\psi_{i0}, \text{sd}(\hat{\xi}_{i0})^2)$.
 In the second scenario, $X=1$ for all patients in site $i$. 
 The same reduced outcome model is fitted among patients with $X=1$. Therefore, $\xi_{i0}$ is the treatment effect for patients with $X=1$ in site $i$, i.e., $\xi_{i0} = \psi_{i0} + \psi_{i1}$. 
In this case, without examining the data one might na\"ively link the estimate $\hat{\xi}_0$ to $\psi_0$ via $\hat{\xi}_0 \sim N(\psi_0, \text{sd}(\hat{\xi}_0)^2)$, but the correct specification is 
    $\hat{\xi}_{i0} \sim N(\psi_{i0}+\psi_{i1}, \text{sd}(\hat{\xi}_{i0})^2)$. 
A second toy example is provided in Web Appendix S2 of the Supplementary Materials. 

\subsubsection{Model sparsity}
Often, many potential tailoring variables are available but only a few are truly predictive of patient response to  treatment. Including all available covariates in the ITR estimation will result in near-zero treatment-covariate interactions, which may result in invalid treatment recommendations and an uninterpretable optimal ITR. To address this, shrinkage priors which aim to shrink small effects towards zero are considered. We use horseshoe priors, though  alternatives exist (see \cite{van2019shrinkage} for a review). Specifically, for treatment-covariate interactions, 
we assume $\psi_t \sim N(0, \tau^2 \lambda_t^2)$, 
where $\lambda_t$ and $\tau$ are local and global shrinkage parameters, respectively, and $\lambda_t, \tau \sim \text{Half-Cauchy}(0,1)$. The shrinkage prior is not placed on  the main effect of treatment, $\psi_0$, as we assume the treatment under consideration has at least some effect on the outcome of interest. We select the treatment-covariate interactions based on posterior credible intervals~\citep{van2019shrinkage}: if $[\psi_{t,0.025},\psi_{t,0.975}]$ does not include zero, the treatment-covariate interaction corresponding to $\psi_t$ will be selected, where $\psi_{t, 0.025}$ and $\psi_{t, 0.975}$ are the $2.5$-th and $97.5$-th percentiles of the posterior distribution of $\psi_t$, respectively.

\section{Simulation studies} \label{sec3}
The simulation study is reported following the scheme proposed in \cite{morris2019using}. A brief 
summary is given below, with details provided in Web Appendix S3 of  
the Supplementary Materials. 

\subsection{Overview}
The simulation study aims to evaluate ITR estimation for a continuous outcome when the individual-level data from multisite studies are not shared across sites, varying: the confounding across sites (both variable set and strength), the degree of heterogeneity across sites, and  the choice of prior distribution used in the analysis. Simulations consider both the case of binary treatments and, inspired by our motivating example  of the International Warfarin Pharmacogenetics Consortium, a continuous dose. We include a sparse data setting (again, mimicking an aspect of the Warfarin data) and explore the use of shrinkage priors in a setting where many  covariates are available, but most are not relevant for optimal treatment decisions. See Web Appendix S3-S4 for details.

Performance is measured via the bias of estimators of the blip parameters relative to their true values, the standard deviation of the estimators, the difference between the value function (dVF) under the true optimal ITR and the value function under the estimated optimal ITR, and the standard deviation of dVF when the estimated treatment rule was applied to the same population. For the many covariates setting, these measures are assessed over: (1) a full set of 2000 iterations, and (2) a subset of iterations where the non-zero treatment-covariate interactions are correctly selected. Results are compared to those obtained from a one-stage analysis based on the full individual-level data.  

\subsection{Results} \label{subsec34}
Here, we present the simulation results for the small sample size and binary treatment under different confounding scenarios, heterogeneity levels, and half-Cauchy (0,1) prior. For brevity, only results for $\psi_0$ and dVF are presented here; we also present the estimates of blip parameters in the sparse data setting with small sample size, different heterogeneity levels and half-Cauchy (0,1) prior, and results for $\psi_1,\psi_2$ and dVF in the many covariates setting; all other results 
are presented in the Supplementary Materials, Web Appendix S5. 

\begin{figure}[!ht]
    \centering
    \includegraphics[width=\textwidth]{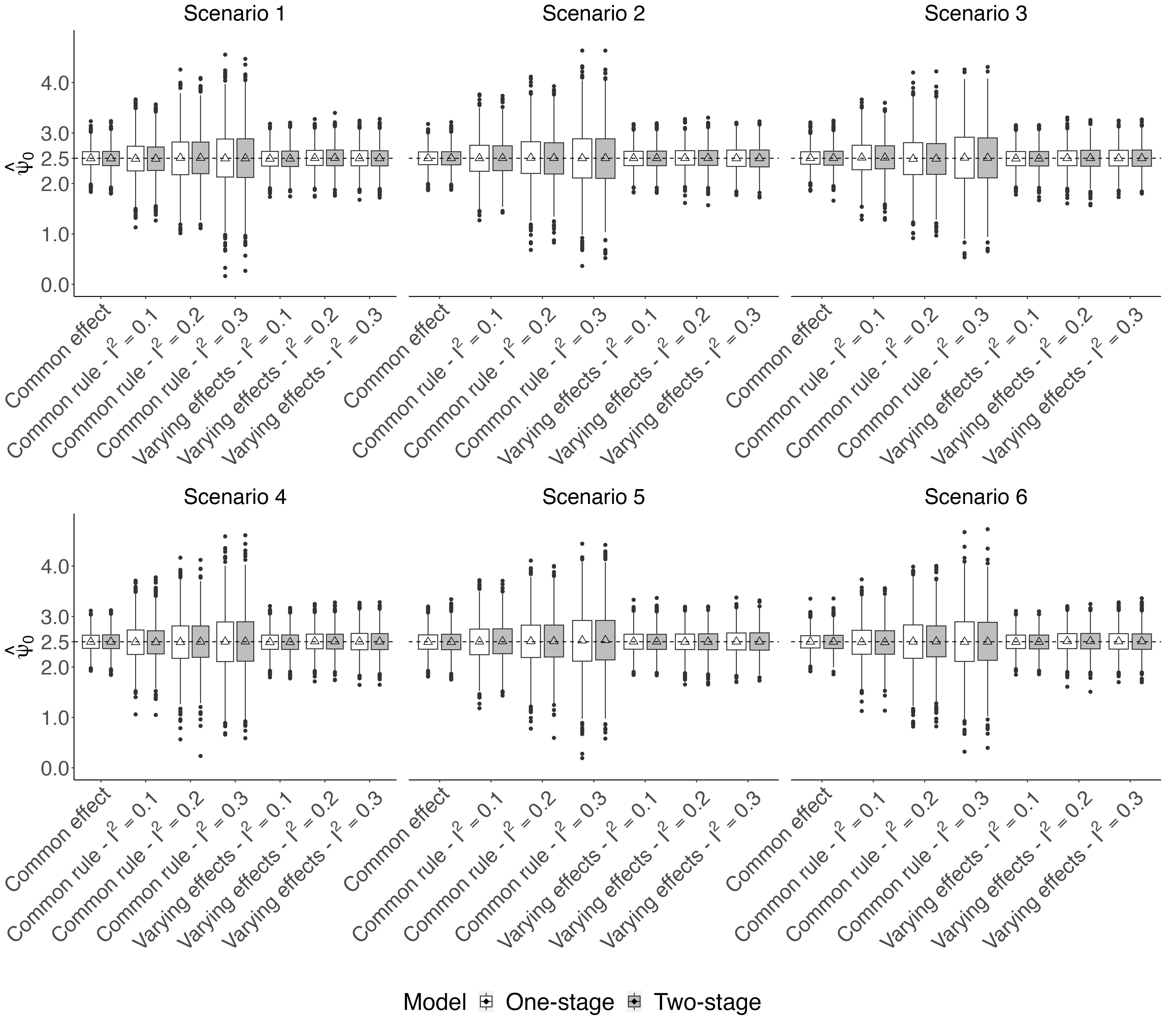}
    \caption{Simulation results for the small sample size and the binary treatment setting. Performance of the methods is assessed over 2000 iterations. Estimates (posterior means) of $\psi_0$ are shown under different confounding scenarios, heterogeneity levels ($I^2 = 0.1, 0.2, 0.3$),  and half-Cauchy (0,1) prior. The triangles represent the mean of the estimates in each case. The dashed line shows the true value of 2.5.}
    \label{fig:n=50-binary-psi0}
\end{figure}

Estimates of $\psi_0$ are presented in Figure \ref{fig:n=50-binary-psi0}. Relative bias is typically less than $1\%$, and neither relative bias nor standard deviation vary much across different confounding scenarios. When the confounding effect is larger, the relative bias is slightly larger. 
Relative bias is similar for different heterogeneity levels, but the variability of the estimators increases with the heterogeneity level. The variation in $\hat{\psi}_0$ in the common rule setting is greater than that in the varying effects setting, a consequence of the data-generating mechanism was such that heterogeneity in $\psi_{i0}$ is greater in the common rule setting than in the varying effects setting. 

\begin{figure}[!ht]
    \centering
    \includegraphics[width=\textwidth]{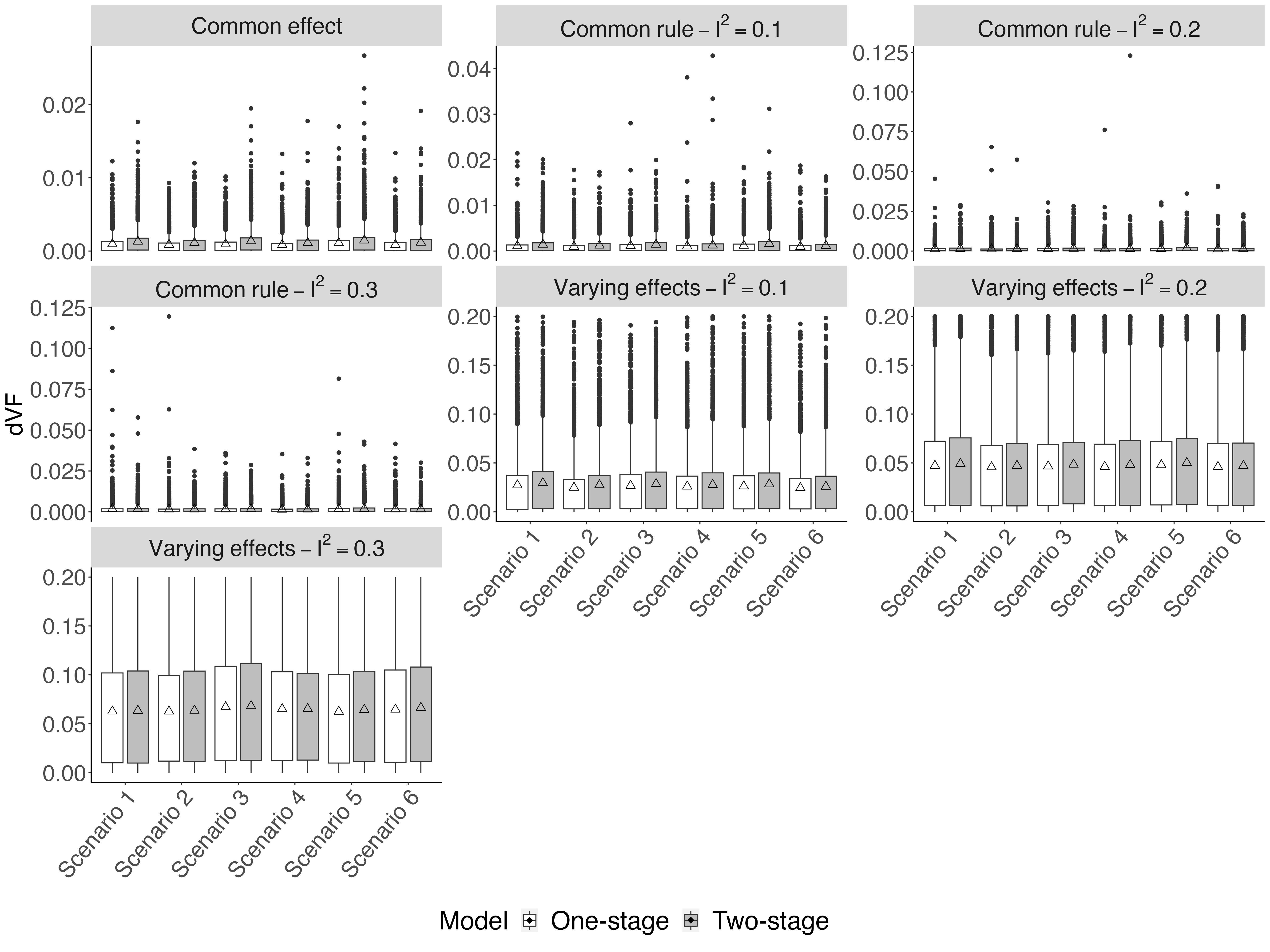}
       \caption{Simulation results for the small sample size and the binary treatment setting. Performance of the methods is assessed over 2000 iterations. The difference in the value function (dVF) between the true and estimated optimal ITR is shown under different confounding scenarios, heterogeneity levels ($I^2 = 0.1, 0.2, 0.3$),  and half-Cauchy (0,1) prior. The triangles represent the mean of the estimates in each case. A smaller dVF corresponds to a better optimal ITR estimation.
    }
    \label{fig:n=50-binary-dvf}
\end{figure}

The dVF 
is shown in Figure \ref{fig:n=50-binary-dvf}. A smaller dVF corresponds to better ITR estimation. The values of the estimated optimal ITR are comparable across different confounding scenarios. When the confounding effect is larger, the dVF is slightly larger.  In the common rule and common effect settings, the dVF is near 0 and varies little. 
The dVF is larger 
with increasing heterogeneity. 
This is unsurprising, since this is a scenario where a single ITR will not provide the optimal treatment for all individuals; rather, the truly optimal treatment is site-specific. However, implementing site-specific rules in a real-world setting is impractical. 

\begin{figure}[!ht]
    \centering
    \includegraphics[width=\textwidth]{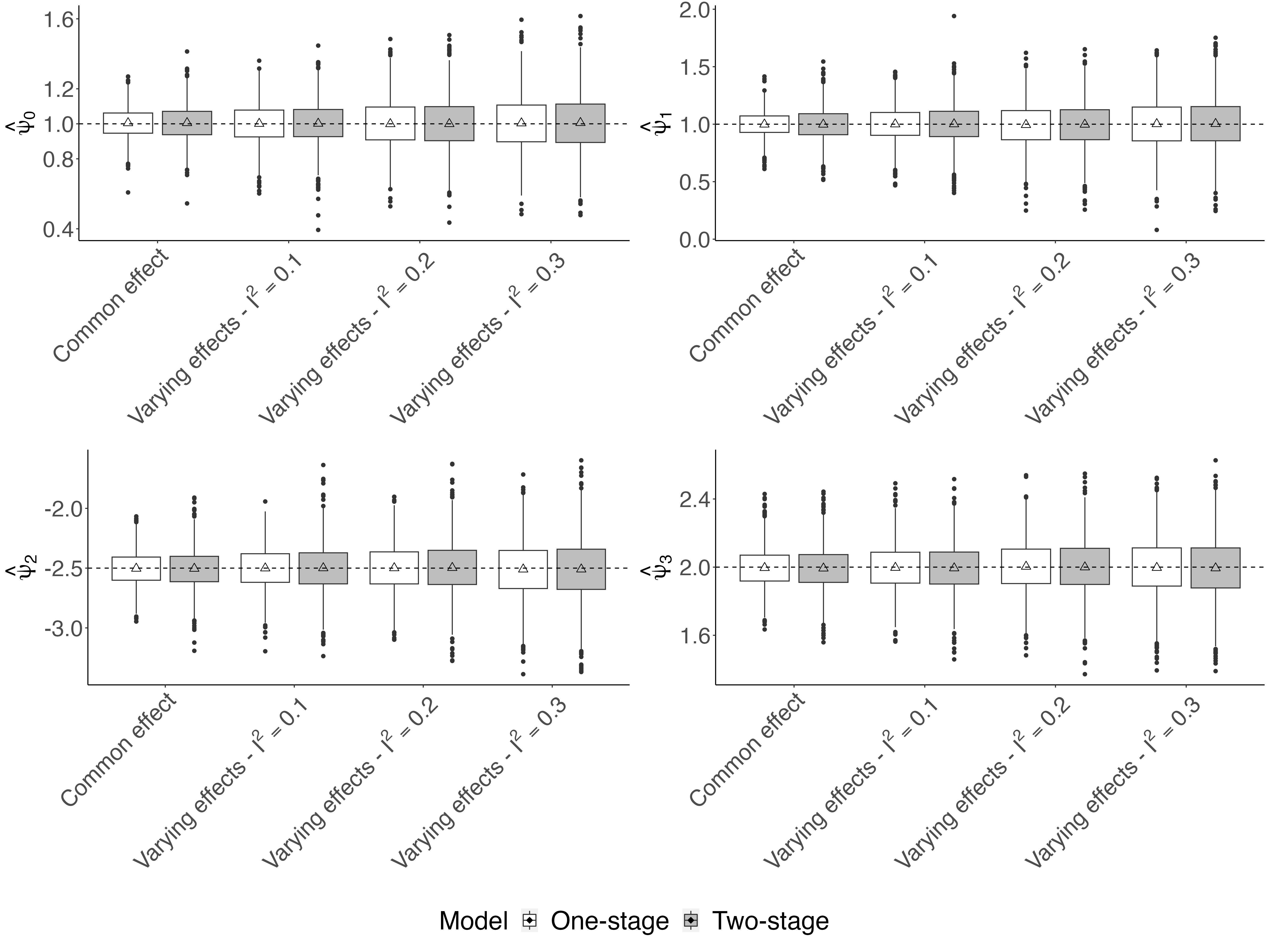}
    \caption{Simulation results for the small sample size and the sparse data setting. Performance of the methods is assessed over 2000 iterations. Estimates (posterior means) of $\psi_0$, $\psi_1$, $\psi_2$, $\psi_3$ are shown under different heterogeneity levels ($I^2 = 0.1, 0.2, 0.3$) and half-Cauchy (0,1) prior. The triangles represent the mean of the estimates in each case. The dashed lines show the true values of $\psi_0$, $\psi_1$, $\psi_2$, $\psi_3$.}
    \label{fig:n=50-sparse}
\end{figure}

Blip parameter estimates in the sparse data setting are shown in Figure \ref{fig:n=50-sparse}. The estimators are unbiased and variability increases with heterogeneity. In the sparse data setting, the dVF (not shown) is zero for all but one or two simulation runs, and it does not change much with different heterogeneity levels. 

Choice of prior made little difference to performance. The model results also do not differ much between the one- and two-stage approaches, but the variability in the two-stage approach is slightly larger than in the one-stage approach when the heterogeneity is small since the two-stage approach cannot share information across sites.

The relative bias and standard deviation of $\hat{\psi}_1$ and $\hat{\psi}_2$, the proportion of selection and the dVF between the true and estimated optimal ITR for the many covariates setting are reported in Table \ref{tab:variable_selection}.  It is more difficult to correctly detect the treatment-covariate interaction when the effect is small or a large number of noisy variables is present. 
When only 10 candidate tailoring variables are considered, non-zero $\psi_1$ is detected in the one- and two-stage models  across 75.4 \% and 71.8\% simulation runs, respectively;  these numbers drop to 49.2\% and 46.5\% when 20 covariates are considered. In all cases, the larger $\psi_2$ is correctly identified for all simulated datasets. Due to the poorer performance in detecting the small non-zero $\psi_1$,  the relative bias of $\hat{\psi}_1$ assessed over  all simulation runs is large; among the simulation iterations when $\psi_1$ is correctly identified, the  relative bias  is $1-2\%$.   Consistent observations can be found for dVF. The horseshoe prior and the selection criterion we adopt may not accurately detect smaller treatment-covariate interactions. However,  if the non-zero effect has been identified, we achieve low bias for parameter estimation and small dVF, corresponding to good ITR estimation.

\begin{sidewaystable}[ht]
\centering
\caption{Simulation results for the many covariates setting. The relative bias (\%, denoted by RB) and  standard deviation (SD) of  $\hat{\psi}_1$ and $\hat{\psi}_2$,  the proportion of selection, and the   difference in the value function (dVF) between the true and estimated optimal ITR  are reported under different numbers of covariates, models, sets of simulation iterations, and a half-Cauchy (0,1) prior. }
\label{tab:variable_selection}
\begin{tabular}{lrrrrrrr}
  \toprule
 Number of covariates & Model & Set & \multicolumn{2}{c}{$\psi_1$} & \multicolumn{2}{c}{$\psi_2$} & dVF (SD) \\ 
 \addlinespace
 \cmidrule{4-5} \cmidrule{6-7}
  & & & RB (SD) & \% Selection & RB (SD) & \% Selection & \\
  \midrule
\multirow{5}{*}{10} & \multirow{2}{*}{One-stage} & Full & -25.703 (0.257) & \multirow{2}{*}{75.400} & -1.503 (0.228) & \multirow{2}{*}{100.000} & 0.135 (0.178)  \\ 
\addlinespace
&  & Selected & -1.463 (0.174) &  & -1.503 (0.228) &  & 0.037 (0.046)  \\ 
\addlinespace
 & \multirow{2}{*}{Two-stage} & Full & -28.748 (0.265) & \multirow{2}{*}{71.800} & -1.549 (0.230) & \multirow{2}{*}{100.000} & 0.153 (0.188)  \\ 
 \addlinespace
  &  & Selected & -0.764 (0.177) &  & -1.549 (0.230) &  & 0.039 (0.048) \\ 
  \addlinespace
\multirow{5}{*}{20} & \multirow{2}{*}{One-stage} & Full & -49.967 (0.275) & \multirow{2}{*}{49.222} & -1.261 (0.237) & \multirow{2}{*}{100.000} & 0.241 (0.210) \\ 
\addlinespace
 &  & Selected & 1.647 (0.171) &  & -1.261 (0.237) &  & 0.032 (0.042)  \\ 
 \addlinespace
  & \multirow{2}{*}{Two-stage} & Full & -52.554 (0.273) & \multirow{2}{*}{46.475} & -1.332 (0.242) & \multirow{2}{*}{100.000} & 0.257 (0.211)  \\ 
  \addlinespace
 &  & Selected & 2.089 (0.171) & & -1.332 (0.242) &  & 0.034 (0.045) \\ 
   \bottomrule
\end{tabular}
\end{sidewaystable}

\section{Estimating an optimal Warfarin dose strategy} \label{sec4}
\subsection{Context and data source} \label{subsec41}
Warfarin is a widely-used oral anticoagulant for thrombosis and thromboembolism treatment and prevention.  
Establishing an optimal Warfarin dose strategy is vital due to the narrow therapeutic window and the large interindividual variability in patients' response to the drug \citep{
international2009estimation
}. The international normalized ratio (INR) is a measure of the time needed for the blood to clot. It should be closely monitored for the safety and effectiveness of Warfarin dosing. 
Many methods have been proposed for finding the optimal dose rule, and clinical factors, demographics, and genetic variability may play an essential role in interindividual variations in the required dose of Warfarin. 
The International Warfarin Pharmacogenetics Consortium
(\citeyear{international2009estimation}) compared several Warfarin dose algorithms and concluded that a pharmacogenetic algorithm in which both genetic and clinical variables are used to inform the appropriate Warfarin dose performs best.  

We utilize the International Warfarin Pharmacogenetics Consortium data and the proposed approaches to data and model sparsity to estimate an optimal individualized Warfarin dose strategy without sharing individual-level data. Following the work of \cite{
schulz2021doubly,danieli2022preserving}, observations with missing values are removed, leading to a sample size of $n=1732$ from 11 different sites. Two of the eleven sites are removed as they include $<$5 patients. 
Therefore, the sample size is reduced to $n=1727$. The final dataset used in the analyses includes several variables: patient age (binned into 9 groups), sex, race, weight and height centered by the site mean,  an indicator for taking amiodarone (an important interacting drug of Warfarin), and VKORC1 and CYP2C9 genotypes, where the latter two variables are genes that may be associated with the interindividual variation in Warfarin dose requirement.
Information on these variables 
is collectively denoted by the vector $\boldsymbol{x}$  
including the leading constant term of one. The stable Warfarin dose and the corresponding INR for each patient are also available in the data.

\subsection{Analyses and results} \label{subsec42}
In accordance with previous work \citep{schulz2021doubly,danieli2022preserving}, the outcome variable is defined as $Y=-\sqrt{\vert 2.5 - \text{INR}\vert}$ such that larger values of $Y$ are clinically preferable. When $Y$ is closer to zero, the INR is closer to the midpoint of the therapeutic window. The observed distribution of the outcomes is roughly symmetric, with values varying from -1.0954 to 0.   
We define a treatment-free function including the main effects of all available variables for the stage-one analysis of the site-specific data.  The blip function is assumed to be quadratic in Warfarin dose, including the main effects of dose, squared-dose, and their interactions with the available variables not including height and weight ($\boldsymbol{x^{(\psi_i^{(1)})}} = \boldsymbol{x^{(\psi_i^{(2)})}} = \boldsymbol{x}/\{\text{weight},\text{height}\}$). Therefore, the outcome model for  site $i$ is 
 $   E(Y_i \vert \boldsymbol{X} = \boldsymbol{x}, A = a) =   \boldsymbol{\beta_i^T} \boldsymbol{x} + a\boldsymbol{\psi_i^{(1)T}x^{(\psi_i^{(1)})}} + a^2\boldsymbol{\psi_i^{(2)T}x^{(\psi_i^{(2)})}}$,
where $a$ is the Warfarin dose, and $\boldsymbol{\beta_i}$ represents the main effects of the available variables on the outcome in site $i$ through the treatment-free function. The site-specific blip parameter vectors $\boldsymbol{\psi_i^{(1)}}=(\psi_{i0}^{(1)}, \ldots, \psi_{i8}^{(1)})$ and $\boldsymbol{\psi_i^{(2)}}=(\psi_{i0}^{(2)}, \ldots, \psi_{i8}^{(2)})$ include main effects of dose and squared-dose (i.e., $\psi_{i0}^{(1)}$ and $\psi_{i0}^{(2)}$), and their interactions with all predictors contained in $\boldsymbol{x^{(\psi_i^{(1)})}} = \boldsymbol{x^{(\psi_i^{(2)})}}$ (i.e., $\psi_{i1}^{(1)}, \ldots, \psi_{i8}^{(1)}$ and $\psi_{i0}^{(2)},\ldots, \psi_{i8}^{(2)}$).

To assess the impact of attempting to preserve individual-level data, the proposed two-stage IPD meta-analysis is implemented. A linear regression is used as the stage-one model to obtain the site-specific blip parameter estimates $\boldsymbol{\hat{\psi}_i^{(1)}}$ and $\boldsymbol{\hat{\psi}_i^{(2)}}$. The site-specific blip parameter estimates were found to be small in magnitude (i.e., close to zero), raising the question of  whether the available covariates have important tailoring effects on the optimal Warfarin dosing in the dataset. Additionally,  in some sites there are not enough patients to estimate the site-specific racial effects, VKORC1 or CYP2C9 genetic effects. Therefore, the Bayesian hierarchical model in the second stage has to be adapted for data sparsity to estimate the common blip parameters $\boldsymbol{\psi^{(1)}}=(\psi_{0}^{(1)}, \ldots, \psi_{8}^{(1)})$ and $\boldsymbol{\psi^{(2)}}=(\psi_{0}^{(2)}, \ldots, \psi_{8}^{(2)})$.
 Horseshoe priors~\citep{shrinkage} are assumed for all treatment-covariate interactions to select variables that truly influence Warfarin dosing by shrinking small effects to zero. Normal priors with mean zero and variance 10,000 are used for the main effects of dose and squared-dose, as Warfarin dose does have effects on the outcome and thus there is no reason to shrink its effects towards zero. The details of the model is described in Web Appendix S6 of the Supplementary Materials.  

The posterior distribution of the optimal dose for an individual with vector $\boldsymbol{x}$ is then approximated by substitution of posterior samples of $\psi_{t}^{(u)}$, $t=0,\ldots, 8$, $u=1,2$ in the maximizer to the common blip function, i.e., 
 $   a^{opt} = -\frac{1}{2}\frac{\boldsymbol{\psi^{(1)T}x^{\psi_i^{(1)}}}}{\boldsymbol{\psi^{(2)T}x^{\psi_i^{(2)}}}}$.
No significant treatment-covariate interactions are selected. Therefore, the optimal dose is fully determined by $\psi_0^{(1)}$ and $\psi_0^{(2)}$, leading to the same recommended dose distribution to all patients and the optimal dose is 41.88 mg/week (posterior median). Not  knowing the true optimal dose in this real-data analysis, we also compare the results of the two-stage approach that avoids disclosing site-specific individual-level data to a one-stage approach that requires all individual-level data to be used at once (and thus requires sharing of data outside the sites at which they were collected).  A consistent conclusion is obtained in the one-stage approach that none of the covariates under consideration have significant tailoring effect on the optimal Warfarin dose, and the posterior median estimate of the common recommended dose is 41.97 mg/week.

\section{Discussion} \label{sec5}
Estimation of optimal ITRs requires the identification of individuals who benefit most from particular treatments and thus the detection of treatment-covariate interactions. Due to the low power associated with detecting the interactions, large datasets may be required to develop an effective optimal ITR, which motivates collaboration across different sites. In multisite studies, the sharing of individual-level data is sometimes constrained, which poses statistical challenges for estimating ITR. In this paper, we adopt a Bayesian two-stage IPD meta-analysis approach to estimate the optimal ITR using multisite data without sharing individual-level data. 

In the presence of treatment-covariate interactions,  traditional meta-regression based on aggregate data are prone to ecological bias and may not reflect the individual-level interactions. However, a two-stage IPD meta-analysis avoids such bias by estimating treatment-covariate interactions within each site separately at the first stage and then synthesising these estimates at the second stage \citep{berlin2002individual, simmonds2007covariate, fisher2011critical}. In our Bayesian two-stage IPD meta-analysis approach, we estimate site-specific optimal ITRs using the regression-based method of Q-learning implemented via linear regression at the first stage, however alternatives such as dWOLS or G-estimation could also have been employed. At the second stage, the site-specific blip parameter estimates are shared to the common analysis center where a Bayesian hierarchical model is used to combine the estimates. We also consider sparsity: estimation of identical models or all parameters of interest at all sites may be infeasible due to heterogeneity of patient populations across sites, and the number of non-zero parameters is often small. We address data sparsity by reparametrizing the likelihood in the second stage such that the site-specific estimates are linked to the correct set of parameters, and use a horseshoe prior and a credible interval selection criterion to select significant treatment-covariate interactions to account for model sparsity. Simulations demonstrate that our approach gives consistent estimation of the common blip  parameters which fully characterize the optimal ITR. When the site-specific optimal ITRs are not very heterogeneous, the value function of the estimated optimal ITR is also close to that of the true optimal ITR.  

We estimate an optimal Warfarin dosing strategy using data from the International Pharmacogenetics Warfarin Consortium. Shrinkage priors are used to select the covariates that truly have effect on the optimal Warfarin dosing. We compare the results obtained from the two-stage approach with the results obtained from a one-stage approach using the full individual-level data. We found that both approaches are consistent in the sense that none of the covariates are selected for the Warfarin dosing strategy, and both provide very close dose recommendations (41.88 vs. 41.97 mg/week).  We also emphasize that although our results may provide some guidance for the establishment of the optimal Warfarin dosing strategy, it is unlikely to provide the whole picture. Several important predictors are not included in the dataset such as alcohol consumption or Vitamin K intake~\citep{international2009estimation},  and thus could not be considered in the analysis.

The simulation results and Warfarin analysis illustrate  our approach's potential to avoid individual-level data sharing when using multisite data to estimate the optimal ITR.  Our approach allows for heterogeneity across sites, which has not been explored in the previous work of \cite{danieli2022preserving, moodie2022privacy} but may be a more reasonable and realistic assumption.
In addition, our approach is quite flexible in the sense that at the first stage we could use any regression-based method to estimate the site-specific blip parameters, as long as the unbiasedness, normality and consistency of the estimators are guaranteed. We focus on simple and interpretable regression, but this approach does require correct specification of the regression model. If the regression model is  misspecified, the site-specific estimators obtained in the first stage could be biased for the true site-specific parameters, leading to biased common parameter estimators (and thus a suboptimal ITR) in the second stage. Semi-parametric alternatives that offer double robustness~\citep{robins2004optimal, wallace2015doubly} could also be considered for stage-one models,  allowing for the  use of more flexible specifications  of the covariate effects (i.e., the treatment-free model, which is essentially a nuisance model).
Additionally, the use of a Bayesian approach in the second stage allows for the seamless incorporation of  the accumulating or external information regarding the optimal ITR from various sources. We may also incorporate variable selection without much additional efforts by using shrinkage priors. This is useful, especially in observational studies where many variables that are irrelevant for the treatment decision may also be collected. In this paper, we use a horseshoe prior and select the treatment-covariate interactions if the $95\%$ posterior credible intervals do not include zero.  The horseshoe prior and credible interval selection criterion are standard in Bayesian variable selection. Simulation shows that our choice may not be powerful enough to detect very small effects.  Other shrinkage priors and alternative selection criteria can also be considered~\citep{
bondell2012consistent,
doi:10.1080/01621459.2014.993077, li2017variable, van2019shrinkage}. 

The proposed approach can be easily adapted for data sparsity, as demonstrated both in simulation and with the Warfarin data. The data sparsity within individual sites considered in this paper only occurs in covariates. The sites in the Warfarin study,  as well as in our simulated examples,  recruited from different (but not entirely distinct) populations, but all treatment choices under investigation were available in all sites.  Sparsity in covariates may restrict the generalization of the estimated site-specific ITRs to a broader population. For example,  without additional assumptions, the estimated site-specific ITRs for sites with only White patients in the samples may not generalize well to the non-White population or a target population including non-White people. However,  this is not, in general, of concern since our estimand of interest is a common rather than site-specific optimal ITR. However, a related concern is that the positivity (or overlap) assumption, often made in the context of causal inference, is violated.  There was, in fact,  a moderate lack of overlap in the Warfarin data (visual inspection of the overlap can be found in Figure S11 in the Web Appendix S6 of the Supplementary Materials).  In the presence of the non-overlap, estimation of parameters of interest requires  extrapolation which may introduce bias.  However, the induced bias might be negligible, if the model is correctly specified and the relationship between treatment, covariates and outcome is consistent across the covariate space,  which is possible in our case. A more concerning issue resulting from the lack of overlap is  the increased variability of parameter estimators,  especially when the estimated effects are all very small (as in this analysis)  leading to a more variable and less reliable estimate of the optimal ITR.  

Although this paper focuses on a continuous outcome, it could be easily extended to outcomes of other types by directly using the existing model of the particular outcome type as the stage-one model~\citep{ 10.2307/27798905,doi:10.1080/01621459.2016.1155993, doi:10.1080/02664763.2017.1386773,simoneau2020estimating}. 
Besides the concerns regarding the individual-level data sharing, 
the two-stage approach is also computationally efficient compared with the one-stage approach using the full individual-level data \citep{burke2017meta}. In our Warfarin analysis, consistent conclusions are obtained in both approaches. However, 
the one-stage approach requires evaluating the likelihood with a large amount of IPD and estimating a large number of parameters simultaneously in a single model,  which increases the computational burden.

Our approach also has some limitations. First, each site is required to have sufficient statistical knowledge of ITR estimation, as site-specific optimal ITRs are estimated separately at each site. This might increase the funding burden for each site in practice. Second, our simulations show that in comparison with the estimated optimal ITR obtained under the scenario of homogeneous optimal ITRs across sites, 
the estimated optimal ITR obtained in the presence of heterogeneity in site-specific optimal ITRs performs worse in terms of the value function. 
A larger number of sites might be needed to account for the heterogeneity in the site-specific optimal ITRs.
However, it may not be realistic to have a large number of sites.  Heterogeneity could arise from the diversity in various aspects across sites such as patient characteristics, treatment delivery, measurements, and study designs~\citep{higgins2019cochrane}. When planning a multisite trial for optimal ITR estimation, one may consider standardizing the research protocol (e.g., eligibility criteria, sampling, and study designs) to reduce extraneous heterogeneity. Standardizing research staff training might also be necessary to ensure that each site follows the common protocol strictly and that the data are measured and collected uniformly~\citep{weinberger2001multisite,noda2006strategies}.

Another approach to aggregate heterogeneous linear ITRs in the binary treatment setting across sites could be a maximin projection learning method proposed by \cite{shi2018maximin}. 
However, this method assumes a common main effect of treatment for which the estimation requires the pooling of all individual-level data \citep{shi2018maximin} and thus cannot be applied in our setting, where we aim to avoid sharing any individual-level records. The extension of the maximin projection learning method to varying main treatment effects across sites and more generalized settings (e.g., continuous treatment) may be of interest to consider in future as an alternative approach to the privacy-preserving estimation of ITRs. 

In addition to the pooling and other approaches to ITR estimation without sharing individual data noted in the introduction, a more general approach with stronger guarantees on privacy that has been pursued in other non-ITR contexts is the use of differentially private algorithms~\citep{dwork2008differential}. Standard linear regression is not differentially private, therefore, our two-stage approach is may not be differentially private, although it does offer some degree since the treatment-free parameter estimates never need to be published or shared under our proposed approach. An interesting avenue of future work is to investigate whether the sharing of only the blip model estimates would pose a violation of differential privacy.

\section*{Acknowledgements}
This work is supported by an award from the Canadian Institutes of Health Research CIHR FDN-16726. EEMM is a Canada Research Chair (Tier 1) in Statistical Methods for Precision Medicine and acknowledges the support of a chercheur de mérite career award from the Fonds de Recherche du Québec, Santé. SG acknowledges support from the Natural Sciences and Engineering Research Council of Canada (NSERC), Canadian Institute for Statistical Sciences (CANSSI) and Fonds de recherche du Québec - Sante (FRQS).


%
  \bibliographystyle{biom} 
 \bibliography{biomtemplate}






\section*{Supporting Information}
 
Web Appendices S1 - S6, referenced in Sections~\ref{sec2} - \ref{sec4}, are available with
this paper at the Biometrics website on Wiley Online
Library.\vspace*{-8pt}







\end{document}


\title{Supplementary materials: Sparse two-stage Bayesian meta-analysis for individualized treatments}

\author[1,*]{Junwei Shen}

\author[1]{Erica E. M. Moodie}

\author[1]{Shirin Golchi}

\affil[1]{Department of Epidemiology, Biostatistics and Occupational Health, McGill University, 2001 McGill College Avenue, Suite 1200 Montreal, QC, H3A 1G1, Canada}
\affil[*]{Corresponding author: Junwei Shen, junwei.shen@mail.mcgill.ca}




\date{}
\maketitle

\begin{appendices}
\makeatletter
\renewcommand \thesection{S\@arabic\c@section}
\renewcommand\thetable{S\@arabic\c@table}
\renewcommand \thefigure{S\@arabic\c@figure}
\makeatother

\section{Link with a one-stage approach}
In this section, we illustrate that, under certain assumptions,  similar estimates of the blip function parameters $\boldsymbol{\psi}$ can be obtained in the proposed two-stage approach and a one-stage approach based on the full individual-level data. As mentioned in the main manuscript, in site $i$, we have the site-specific outcome model:
\begin{align*}
    E(Y_{ij} \vert \boldsymbol{X} =\boldsymbol{x_{ij}}, A = a_{ij}) = \boldsymbol{\beta_i^Tx_{ij}^{(\beta)}} + a_{ij} \boldsymbol{\psi_i^T x_{ij}^{(\psi)}},
\end{align*}
where $i \in \{1, \ldots, K\}$ and $j \in \{1, \ldots, n_i\}$ index the site and individual patient in a given site respectively, and $n_i$ is the number of patients in site $i$.
The predictive and prescriptive covariate vectors are denoted by $\boldsymbol{x_{ij}^{(\beta)}}$ and $\boldsymbol{x_{ij}^{(\psi)}}$, respectively.
The $p$-dimensional site-specific treatment-free function parameter and $q$-dimensional blip function parameter are denoted by $\boldsymbol{\beta_i}=(\beta_{i0},\ldots,\beta_{i,p-1})$ and  $\boldsymbol{\psi_i}=(\psi_{i0},\ldots, \psi_{i,q-1})$, respectively. Then, with site-specific estimates $\hat{\psi}_{it}$ and the associated standard deviations sd$(\hat{\psi}_{it})$, for $t = 0, \ldots, q-1,$  obtained from the stage-one models, a Bayesian hierarchical model is implemented in the second stage:
\begin{align*}
\begin{split}
     \hat{\psi}_{it} &\sim N(\psi_{it}, sd(\hat{\psi}_{it})^2), \\
     \psi_{it} & \sim N(\psi_t, \sigma_{\psi_t}^2),\\
     \psi_t & \sim p_{\psi_t}(\psi_t),\\
     \sigma_{\psi_t}^2 &\sim p_{\sigma_{\psi_t}^2}(\sigma_{\psi_t}^2),
     \end{split}
\end{align*}
where $\psi_{it}$ and $\psi_t$ are the $(t+1)$-th elements of the site-specific and common blip function parameter vectors. The between-site heterogeneity associated with $\psi_{it}$ is denoted by $\sigma_{\psi_t}^2$. Prior distributions  $p_{\psi_t}$ and $p_{\sigma_{\psi_t}^2}$ can be assigned for the unknown parameters $\psi_t$ and $\sigma_{\psi_t}^2$. 
The joint posterior distribution for the two-stage approach is then 
\begin{align*}
     &p(\boldsymbol{\psi}, \boldsymbol{\psi_{1}},\ldots, \boldsymbol{\psi_K},  \boldsymbol{\sigma_\psi^2} \vert \boldsymbol{\hat{\psi}_i}, \boldsymbol{\textbf{var}(\hat{\psi}_i)})
    \propto & 
    \underbrace{\prod_{i=1}^K \prod_{t=0}^{q-1} p(\hat{\psi}_{it} \vert \psi_{it}, \text{var}(\hat{\psi}_{it}) )}_{\text{Likelihood}}
    \times 
     \underbrace{ \prod_{i=1}^K \prod_{t=0}^{q-1} p(\psi_{it} \vert \psi_t, \sigma_{\psi_t}^2) p(\boldsymbol{\psi},\boldsymbol{\sigma_\psi^2)}
     }_{\text{Prior}},
\end{align*}
where $\boldsymbol{\hat{\psi}_i}=(\hat{\psi}_{i0},\ldots,\hat{\psi}_{i,q-1})$, $\boldsymbol{\textbf{var}(\hat{\psi}_i)}=(\text{var}(\hat{\psi}_{i0}),\ldots, \text{var}(\hat{\psi}_{i,q-1}))$, and $p(\boldsymbol{\psi},\boldsymbol{\sigma_\psi^2)} = \prod_{t=0}^{q-1} p(\psi_t)\prod_{t=0}^{q-1}p(\sigma_{\psi_t}^2)$.

With the full individual-level data, a one-stage model can be implemented:
\begin{align*}
\begin{split}
    Y_{ij} &= \boldsymbol{\beta_i^T} \boldsymbol{x_{ij}^{(\beta)}} + a_{ij} \boldsymbol{\psi_i^T x_{ij}^{(\psi)}} + \epsilon_{ij},\\
    & = \sum_{s=0}^{p-1} \beta_{is}x_{ijs}^{(\beta)} + a_{ij}\sum_{t=0}^{q-1}\psi_{it}x_{ijt}^{(\psi)} + \epsilon_{ij},
    \end{split}
\end{align*}
where the residual error $\epsilon_{ij}$ follows a normal distribution with mean 0 and within-site residual variance $\sigma_i^2$. The site-specific parameters $\beta_{is}$, $\psi_{it}$ for $i=1,\ldots,K$, $s=0,\ldots,p-1$, $t=0,\ldots,q-1$  
satisfy 
\begin{align}
\label{eq3}
\begin{split}
   \beta_{is}&\sim N(\beta_s, \sigma_{\beta_s}^2), \\
   \psi_{it} &\sim N(\psi_t, \sigma_{\psi_t}^2),
\end{split}
\end{align}
where $\boldsymbol{\beta} = (\beta_0,\ldots, \beta_{p-1})$ and $\boldsymbol{\psi}=(\psi_0,\ldots,\psi_{q-1})$ are the common treatment-free and blip function parameters, respectively. 
We note that the distributional assumption (\ref{eq3}) is slightly different from the distributional assumption 
\begin{align}
    \boldsymbol{\psi_i} \sim MVN(\boldsymbol{\psi}, \boldsymbol{\Sigma_\psi}).
    \label{eq2}
\end{align}
As discussed in the main manuscript, in the two-stage approach, the site-specific treatment-free function parameter estimates are ignored in the second stage. Therefore, only assumption (\ref{eq2}) is required to pool the blip function parameter estimates, although assumption (\ref{eq3}) is also reasonable. In the Bayesian framework, priors will be assigned to the unknown parameters $\beta_s$, $\psi_t$, $\sigma_i^2$, $\sigma_{\beta_s}^2$, $\sigma_{\psi_t}^2$, $i=1,\ldots,K,s=0,\ldots,p-1,t=0,\ldots,q-1$:
\begin{align*}
\beta_{s} &\sim  p_{\beta_s}(\beta_s), &
\psi_t & \sim p_{\psi_t}(\psi_t),  & & \\
 \sigma_i^2 &\sim p_{\sigma_i^2}(\sigma_i^2), &
  \sigma_{\beta_s}^2 &\sim p_{\sigma_{\beta_s}^2}(\sigma_{\beta_s}^2),  &
    \sigma_{\psi_t}^2 &\sim p_{\sigma_{\psi_t}^2}(\sigma_{\psi_t}^2).
\end{align*}

Therefore, the joint posterior distribution for the one-stage approach is 

\begin{align*}
      & p( \boldsymbol{\beta}, \boldsymbol{\psi}, \boldsymbol{\beta_{1}}, \ldots, \boldsymbol{\beta_K}, \boldsymbol{\psi_1}, \ldots, \boldsymbol{\psi_K},  \boldsymbol{\sigma^2}, \boldsymbol{\sigma_\beta^2},  \boldsymbol{\sigma_\psi^2} \vert \boldsymbol{Y_1}, \ldots, \boldsymbol{Y_K} )\\
    \propto & \underbrace{
    \prod_{i=1}^K \prod_{j=1}^{n_i} p(Y_{ij} \vert \boldsymbol{\beta_i}, \boldsymbol{\psi_i},\sigma_i^2) 
    }_{\text{Likelihood}} \times  
     \underbrace{\prod_{i=1}^K \prod_{s=0}^{p-1} p(\beta_{is} \vert \beta_s, \sigma_{\beta_s}^2) \prod_{i=1}^K \prod_{t=0}^{q-1} p(\psi_{it} \vert \psi_t, \sigma_{\psi_t}^2)p(\boldsymbol{\beta},\boldsymbol{\psi},\boldsymbol{\sigma^2},\boldsymbol{\sigma_\beta^2},\boldsymbol{\sigma_\psi^2})}_{\text{Prior}},
\end{align*}
where $\boldsymbol{Y_i}=(Y_{i1},\ldots,Y_{i,n_i})$, $\boldsymbol{\sigma^2}=(\sigma_1^2,\ldots,\sigma_K^2)$, $\boldsymbol{\sigma_\beta^2}=(\sigma_{\beta_0}^2,\ldots,\sigma_{\beta_{p-1}}^2)$, $\boldsymbol{\sigma_{\psi}^2}=(\sigma_{\psi_0}^2,\ldots, \sigma_{\psi_{q-1}}^2)$, and independent priors can be assigned to $\beta_s$, $\psi_t$, $\sigma_i^2$, $\sigma_{\beta_s}^2$, $\sigma_{\psi_t}^2$ such that $p(\boldsymbol{\beta},\boldsymbol{\psi},\boldsymbol{\sigma^2},\boldsymbol{\sigma_\beta^2},\boldsymbol{\sigma_\psi^2}) = \prod_{s=0}^{p-1} p(\beta_s)\prod_{t=0}^{q-1} p(\psi_t)\prod_{i=1}^K p(\sigma_i^2)\prod_{s=0}^{p-1} p(\sigma_{\beta_s}^2) \prod_{t=0}^{q-1}p(\sigma_{\psi_t}^2)$. Thus, all parameters are estimated at once in the one-stage approach, while only blip function parameters and their related between-site variances are estimated separately in the two-stage approach. 
To see the similarity between the two approaches, we show that, under certain assumptions, $\prod_{j=1}^{n_i}p(Y_{ij}\vert \boldsymbol{\beta_{i}}, \boldsymbol{\psi_{i}}, \sigma_i^2)$ and $p(\hat{\psi}_{it}\vert \psi_{it}, \text{var}(\hat{\psi}_{it}))$ carry the same information of $\psi_{it}$. Define $Y_{ijt} = Y_{ij} - \sum_{s=0}^{p-1} \beta_{is}x_{ijs}^{(\beta)} - a_{ij} \sum_{t'\neq t}\psi_{it'}x_{ijt'}^{(\psi)}$ and $\Tilde{Y}_{ijt}=Y_{ij} -  \sum_{s=0}^{p-1} \hat{\beta}_{is}x_{ijs}^{(\beta)} - a_{ij} \sum_{t'\neq t}\hat{\psi}_{it'}x_{ijt'}^{(\psi)}$. 
Without loss of generality, assume that the focus now is only on $\psi_{t_0}$, $\psi_{it_0}$, and $\sigma_{\psi_{t_0}}^2$ for some $t_0 \in \{0,\ldots,q-1\}$, and other parameters (e.g., $\sigma_i^2$, $\beta_{is}$, $\psi_{it}$, $t\neq t_0$) are nuisance parameters. The likelihood in the one-stage approach
\begin{align*}
\begin{split}
    &\prod_{j=1}^{n_i} p(Y_{ij} \vert \boldsymbol{\beta_i}, \boldsymbol{\psi_i}, \sigma_i^2)
    \propto  \exp{ \left\{-\frac{1}{2\sigma_i^2} \sum_{j=1}^{n_i} (Y_{ijt_0} - a_{ij} \psi_{it_0}x_{ijt_0}^{(\psi)})^2 \right\}}\\
    \propto & \exp \left\{-\frac{1}{2\sigma_i^2}\left (\psi_{it_0}^2\sum_{j=1}^{n_i} a_{ij}^2 (x_{ijt_0}^{(\psi)})^2-2\psi_{it_0}\sum_{j=1}^{n_i} a_{ij}x_{ijt_0}^{(\psi)} Y_{ijt_0}\right)\right\} .
\end{split}
\end{align*}
The likelihood in the two-stage approach 
\begin{align*}
\begin{split}
    p(\hat{\psi}_{it_0} \vert \psi_{it_0}, \text{var}(\hat{\psi}_{it_0})) &\propto \exp \left\{ -\frac{1}{2\text{var}(\hat{\psi}_{it_0})} (\hat{\psi}_{it_0} - \psi_{it_0})^2\right\} \\
    & \propto \exp \left\{-\frac{1}{2\text{var}(\hat{\psi}_{it_0})} (\psi_{it_0}^2 - 2\hat{\psi}_{it_0} \psi_{it_0}) \right\}\\
     &  \propto \exp \left\{-\frac{\psi_{it_0}^2 \sum_{j=1}^{n_i} a_{ij}^2 (x_{ijt_0}^{(\psi)})^2 - 2\psi_{it_0}\sum_{j=1}^{n_i} a_{ij}x_{ijt_0}^{(\psi)} \Tilde{Y}_{ijt_0}}{2\text{var}(\hat{\psi}_{it_0})\sum_{j=1}^{n_i} a_{ij}^2 (x_{ijt_0}^{(\psi)})^2} \right\},
\end{split}
\end{align*}
since the ordinary least squares (OLS) estimator is given by 
\begin{align*}
    \hat{\psi}_{it_0} = \frac{\sum_{j=1}^{n_i} \Tilde{Y}_{ijt_0} a_{ij} x_{ijt_0}^{(\psi)}}{\sum_{j=1}^{n_i} a_{ij}^2(x_{ijt_0}^{(\psi)})^2}.
\end{align*}
When $\hat{\beta}_{is} = \beta_{is}$, and $\hat{\psi}_{it} = \psi_{it}$ for $s=0,\ldots, p-1$, $t\neq t_0$, that is, $\beta_{is}$ and $\psi_{it}$ are estimated with negligible error in the site-specific linear regression model, then $\Tilde{Y}_{ijt_0}=Y_{ijt_0}$ and $\text{var}(\hat{\psi}_{it_0}) = \frac{\sigma_i^2}{\sum_{j=1}^{n_i} a_{ij}^2 (x_{ijt_0}^{(\psi)})^2}.$ Thus, we have 
\begin{align*}
     p(\hat{\psi}_{it_0} \vert \psi_{it_0}, \text{var}(\hat{\psi}_{it_0})) \propto \exp \left\{-\frac{1}{2\sigma_i^2}\left (\psi_{it_0}^2\sum_{j=1}^{n_i} a_{ij}^2 (x_{ijt_0}^{(\psi)})^2-2\psi_{it_0}\sum_{j=1}^{n_i} a_{ij}x_{ijt_0}^{(\psi)} Y_{ijt_0}\right)\right\}, 
\end{align*}
and  $p(\hat{\psi}_{it_0} \vert \psi_{it_0}, \text{var}(\hat{\psi}_{it_0}))$ contains the same information of $\psi_{it_0}$ as $\prod_{j=1}^{n_i} p(Y_{ij} \vert \boldsymbol{\beta_{i}}, \boldsymbol{\psi_{i}}, \sigma_i^2)$. This applies to all sites under the assumption that $\beta_{is}$,and $\psi_{it}$
are estimated with negligible error (i.e., $\hat{\beta}_{is} = \beta_{is}$, $\hat{\psi}_{it} = \psi_{it}$) in the stage-one linear regression models. This assumption can be feasible and approximately true for a moderate to large sample size, given the unbiasedness and consistency of the OLS estimators. Then, with the same common distribution for $\psi_{it_0},i=1,\ldots,K,$ and the same priors for $\psi_{t_0}$ and $\sigma_{\psi_{t_0}}^2$, the posterior distribution of $\psi_{t_0}$, $\boldsymbol{\psi_{(t_0)}}=(\psi_{1,t_0},\ldots,\psi_{K,t_0})$, $\sigma_{\psi_{t_0}}^2$ conditional on $\boldsymbol{Y_i}$, $\boldsymbol{\beta_i}$, $\boldsymbol{\sigma^2}$, $\boldsymbol{\psi_{i(-t_0)}}=\boldsymbol{\psi_i}/\{\psi_{i,t_0}\}$, for $i=1,\ldots,K$,   in the one-stage approach
\begin{align*}
    &p(\psi_{t_0}, \boldsymbol{\psi_{(t_0)}}, \sigma_{\psi_{t_0}}^2 \vert \boldsymbol{Y_i}, \boldsymbol{\beta_i}, \boldsymbol{\psi_{i(-t_0)}}, \boldsymbol{\sigma^2})\\
     \propto & \prod_{i=1}^K \prod_{j=1}^{n_i} p(Y_{ij} \vert \boldsymbol{\beta_i},\boldsymbol{\psi_i},\sigma_i^2) \prod_{i=1}^K p(\psi_{it_0} \vert \psi_{t_0}, \sigma_{\psi_{t_0}}^2)p(\psi_{t_0})p(\sigma_{\psi_{t_0}}^2),
\end{align*}
is equivalent to the joint posterior distribution of $\psi_{t_0},\boldsymbol{\psi_{(t_0)}}, \sigma_{\psi_{t_0}}^2$ given $\boldsymbol{\hat{\psi}_{(t_0)}}=(\hat{\psi}_{1,t_0},\ldots,\hat{\psi}_{K,t_0})$ and $\boldsymbol{\textbf{var}(\hat{\psi}_{(t_0)})}=(\text{var}(\hat{\psi}_{1,t_0}),\ldots,\text{var}(\hat{\psi}_{K,t_0}))$ in the two-stage approach:
\begin{align*}
       &p(\psi_{t_0}, \boldsymbol{\psi_{(t_0)}}, \sigma_{\psi_{t_0}}^2 \vert \boldsymbol{\hat{\psi}_{(t_0)}}, \boldsymbol{\textbf{var}(\hat{\psi}_{(t_0)})})\\
     \propto & \prod_{i=1}^K p(\hat{\psi}_{it_0} \vert  \psi_{it_0}, \text{var}(\hat{\psi}_{it_0}))\prod_{i=1}^K p(\psi_{it_0} \vert \psi_{t_0}, \sigma_{\psi_{t_0}}^2)p(\psi_{t_0})p(\sigma_{\psi_{t_0}}^2),
\end{align*}
which leads to similar estimates in the two approaches.

\section{Data sparsity: A second toy example} 
A simply toy example is provided in the main text. Here, in a second example, we assume one categorical covariate $X$ consisting of three levels (i.e., $p=q=3$) and the true outcome model for an individual at site $i$ is 
      $E(Y\vert X) = \beta_{i0} + \beta_{i1} X_2 + \beta_{i2} X_3+ A(\psi_{i0} + \psi_{i1} X_2 + \psi_{i2}X_3)$.
Here, we choose the first category as the reference, and two indicators $X_2$, $X_3$ are created for the second and third categories. Therefore, $\psi_{i0}$ is the treatment effect for patients in the first category in site $i$; $\psi_{i0} + \psi_{i1}$ is the treatment effect for patients in the second category in site $i$; and $\psi_{10} + \psi_{i1}$ is the treatment effect for patients in the third category in site $i$. When (i) all patients in site $i$ have a covariate value that is in the same category, or (ii) none take the second (or the third) category, but there are patients in the first and the third (or the second) categories, the situations are similar to the first example, and we do not duplicate the discussion. We consider a different scenario where none lie in the reference category, but both the second and third categories are represented in the samples. In this case, one of the last two categories will automatically become the ``new'' reference. Without loss of generality, assume the second category as the new reference. The site-specific outcome model then becomes
    $E(Y \vert X) = \gamma_{i0} + \gamma_{i2}X_3 + A(\xi_{i0} + \xi_{i2} X_3)$,
where $\xi_{i0}$ is the treatment effect for patients in the second category in site $i$, i.e., $\xi_{i0} = \psi_{i0} + \psi_{i1}$; $\xi_{i2}$ is the difference in treatment effects for patients between the third and the second categories in site $i$, i.e., $\xi_{i2} = \psi_{i2}-\psi_{i1}$. Then the likelihood contribution of site $i$ becomes
    $\hat{\gamma}_{i0} \sim N(\psi_{i0} + \psi_{i1}, \text{sd}(\hat{\gamma}_{i0})^2), \quad  \hat{\gamma}_{i2} \sim N(\psi_{i2} - \psi_{i1}, \text{sd}(\hat{\gamma}_{i2})^2)$.
Therefore, it is essential to examine the data \textit{in each site} to detect any cases of sparsity in variable levels before incorporating the site-specific estimates into the model. For each site with sparse data, we update the likelihood contribution in the Bayesian hierarchical model  based on the impact of data sparsity on the model parameter interpretation. Then, priors can be assigned to the common mean parameters and variance component parameters as is shown in the main text. 

\section{Simulation studies: ADEMP reporting} 
\subsection{Aims} \label{subsec31}
The aim of the simulation study is to evaluate ITR estimation for a continuous outcome when the individual-level data from multisite studies is protected from release via a two-stage IPD meta-analysis, under assumptions concerning (1) the confounder sets across sites, (2) the strength of confounding, (3) the degree of heterogeneity across sites, and (4) the choice of prior distribution. Points (1) - (3) concerns the data-generating mechanisms, while (4) concerns the analysis model.

\subsection{Data-generating mechanisms} \label{subsec32}
In the simulation, we primarily focus on the binary treatment setting but include a reduced set of scenarios for the continuous treatment to illustrate the use of the proposed approach in a dosing setting. We also include a sparse data setting which mimics a particular, challenging feature of the International Warfarin Pharmacogenetics Consortium data: not all parameters can be estimated at all sites due to differences in populations across sites.  Additionally, a small simulation is conducted to explore the use of shrinkage priors when a number of covariates are available, but only some are truly relevant for optimal treatment decisions. For all settings (except for the simulations with many covariates) , $K = 10$ sites with an average sample size of $n=50$ (small sample size) or $200$ (large sample size) are assumed for all scenarios. The site-specific sample sizes vary between $0.6n$ and $1.4n$.  For simulations with shrinkage priors in the many covariates setting, only a large sample size is assumed. 

\subsubsection{Binary treatment setting}
Two covariates $X_1, X_2$ are considered and their distributions vary across sites: for sites 3, 6, and 9, $X_1 \sim N(5, 1), X_2 \sim  \text{Bernoulli}(0.5)$; for sites 1, 4, 7, and 10, $X_1 \sim 6\text{Beta}(4,4)+2, X_2 \sim \text{Bernoulli}(0.3) $; for sites 2, 5, and 8, $X_1 \sim \text{U}[2,8], X_2 \sim  \text{Bernoulli}(0.7) $.

The treatment assignment $A$ follows a Bernoulli distribution with the propensity score $P_i(\boldsymbol{x})$ at site $i$ determined by 
    $P_i(A=1\vert \boldsymbol{X} = \boldsymbol{x})=\left [1+e^{-(\alpha_{i0}+\alpha_{i1}x_1+\alpha_{i2}x_2)} \right]^{-1}$,
where $(\alpha_{i0}, \alpha_{i1}, \alpha_{i2})$ for different confounding scenarios are given in Table \ref{confound}. In scenarios 1 and 2, propensity score models are identical across sites, and the confounding effect can be either large (scenario 1) or small (scenario 2). In scenarios 3 and 4, site-specific propensity score models with the same set of confounders are assumed for each site, and two different confounding effects are also assumed. In scenarios 5 and 6, both propensity score model parameters and confounder sets are different across sites.
\begin{table}[!ht]
    \centering
\caption{Parameters in the propensity score model for binary treatment simulations in different scenarios}
 \begin{tabular}{lcc}
    \toprule
  &  Scenario 1 & Scenario 2\\
    \midrule
   $\alpha_{i0}$ &  0.1  & 0.01  \\
   $\alpha_{i1}$ &  0.1  & 0.01\\
   $\alpha_{i2}$ & 0.1 & 0.01\\
   \midrule
   & Scenario 3 & Scenario 4\\
   \midrule
     $\alpha_{i0}$ &  U[0.06, 0.14]  & U[0.006, 0.014]  \\
   $\alpha_{i1}$ &  U[0.06, 0.14]  & U[0.006, 0.014]\\
   $\alpha_{i2}$ & U[0.06, 0.14] & U[0.006, 0.014]\\
   \midrule 
   & Scenario 5 & Scenario 6\\
   \midrule
     $\alpha_{i0}$ &  
      U[0.3,0.7]    & U[0.03, 0.07]  \\
   $\alpha_{i1}$ &  $\begin{cases} 
      0 & i = 1, 3, 5, 7, 9 \\
      \text{U}[0.06,0.14] & i = 2, 4, 6, 8, 10
   \end{cases}$   & $\begin{cases} 
      0 & i = 1, 3, 5, 7, 9 \\
      \text{U}[0.006,0.014] & i = 2,4, 6, 8, 10
   \end{cases}$\\
   $\alpha_{i2}$ & $\begin{cases} 
      \text{U}[0.3,0.7] & i = 1, 3, 5, 7, 9\\
      0 & i=2, 4, 6, 8, 10
   \end{cases}$  & $\begin{cases} 
      \text{U}[0.03,0.07] & i = 1, 3, 5, 7, 9 \\
      0 & i=2,4,6,8,10
   \end{cases}$\\
   \bottomrule
       \label{confound}
    \end{tabular}
\end{table}

Suppressing the individual-specific subscript, the continuous outcome for an individual at site $i$ is generated by 
    $Y_i = \beta_{i0} + \beta_{i1} x_1 + \beta_{i2} x_2 + a(\psi_{i0} + \psi_{i1} x_1) + \epsilon$,
where 
the random error $\epsilon$ follows a normal distribution with mean zero and residual variance $\sigma_\epsilon^2 = 0.25$. 
For the site-specific parameters $\boldsymbol{\theta_i} = (\beta_{i0}, \beta_{i1}, \beta_{i2}, \psi_{i0}, \psi_{i1})$, we consider three different scenarios: 
\begin{itemize}
    \item[--] common effect: $\boldsymbol{\theta_1} = \boldsymbol{\theta_2} = \ldots = \boldsymbol{\theta_{10}}=\boldsymbol{\theta}$ and $\boldsymbol{\theta}=(\beta_0, \beta_1, \beta_2, \psi_0, \psi_1)$ is the common population parameter;
    \item[--] common rule: 
    $\beta_{it} \sim N(\beta_t, \sigma_B^2)$, $\psi_{i1} \sim N(\psi_1, \sigma_B^2)$, for $t=0,1,2$, $i=1, \ldots,10$
     and $\psi_{10}/\psi_{11} = \psi_{20}/\psi_{21} = \ldots = \psi_{10,0}/ \psi_{10,1} = -5$,
  where the between-site variance $\sigma_B^2$ is derived from heterogeneity level $I^2 = \frac{\sigma_B^2}{\sigma_B^2 + \sigma_\epsilon^2} = 0.1, 0.2, 0.3$; 
    \item[--] varying effects:  $ \boldsymbol{\theta_{i}} \sim MVN(\boldsymbol{\theta}, \boldsymbol{\Sigma_{\theta}})$, where $\boldsymbol{\Sigma_\theta}$ is a $5\times 5$ diagonal matrix where the between-site variance is derived from $I^2$ as in the common rule setting.
\end{itemize}

In all three scenarios, the common treatment-free parameters are $\beta_0 = 4$, $\beta_1 =1 $, $\beta_2 =1$ and the common blip parameters are $\psi_0 = 2.5$, $\psi_1 =-0.5$ such that the common optimal ITR is given by $d^{opt}(\boldsymbol{x}) =I(\psi_0+\psi_1 x_1 >0) = I(x_1 <5)$.
The common effect setting assumes that all site-specific parameters are equal to the common population parameters as in the simulation studies in \cite{danieli2022preserving, moodie2022privacy}. No heterogeneity exists in the site-specific blip parameters $\psi_{i0}$ and $\psi_{i1}$ 
and the site-specific optimal ITRs are $d_i^{opt}(\boldsymbol{x}) = I(\psi_{i0} + \psi_{i1}x_1>0)$. The varying effects setting assumes a common multivariate normal distribution for the site-specific parameters. The two blip parameters $\psi_{i0}$ and $\psi_{i1}$ are freely varying across sites. Therefore, heterogeneity exists in $(\psi_{i0}, \psi_{i1})$ and $d_{i}^{opt}(\boldsymbol{x})$. The common rule setting considers heterogeneity scenarios that can be viewed as intermediate between common effect and varying effects; the blip parameters $\psi_{i0}, \psi_{i1}$ are varying across sites, however, the site-specific optimal ITRs $d_i^{opt}(\boldsymbol{x})$ are fixed by restricting the ratio $\psi_{i0}/\psi_{i1}$ to be identical across sites. In this setting, heterogeneity only exists in the blip parameters but not the site-specific optimal ITRs.

\subsubsection{Continuous treatment setting}

For the continuous treatment setting, the same covariates $X_1$, $X_2$ are generated in the same way as the binary treatment setting. The treatment $A \sim N(X_1, 1)$. The outcome for an individual at site $i$ is generated by 
  $ Y_i = \beta_{i0} + \beta_{i1} x_1 + \beta_{i2} x_2 + a(\psi_{i0} + \psi_{i1} a + \psi_{i2} x_1) + \epsilon$,
where the random error $\epsilon$ follows a normal distribution with mean zero and residual variance $\sigma_\epsilon^2 = 0.25$. Two different settings are considered for the site-specific parameters $\boldsymbol{\theta_i} = (\beta_{i0}, \beta_{i1}, \beta_{i2}, \psi_{i0}, \psi_{i1}, \psi_{i2})$:
\begin{itemize}
    \item[--] common effect: $\boldsymbol{\theta_1} = \boldsymbol{\theta_2} = \ldots = \boldsymbol{\theta_{10}}=\boldsymbol{\theta}$ and $\boldsymbol{\theta}=(\beta_0, \beta_1, \beta_2, \psi_0, \psi_1, \psi_2)$ is the common population parameter;
    \item[--] varying effects:  $ \boldsymbol{\theta_{i}} \sim MVN(\boldsymbol{\theta}, \boldsymbol{\Sigma_{\theta}})$, where $\boldsymbol{\Sigma_\theta}$ is a $6\times 6$ diagonal matrix where the between-site variance is obtained from the heterogeneity level $I^2 = \frac{\sigma_B^2}{\sigma_B^2 + \sigma_\epsilon^2} = 0.1, 0.2, 0.3$.
\end{itemize}

In both settings, the common treatment-free parameters are $\beta_0 = 4$, $\beta_1 =1 $, $\beta_2 =1$,  and the common blip parameters are $\psi_0 = 1$, $\psi_1 =-2$, $\psi_2 = 1$. The common optimal ITR is $d^{opt}(\boldsymbol{x}) = \arg \text{max}_a (-2a^2 + a + ax_1) = (1+x_1)/4$.

\subsubsection{Sparse data setting}
As discussed, it is possible in multisite studies that the site-specific parameters cannot be estimated due to an insufficient number of patients with a given set of characteristics. 
To show how the proposed method deals with this scenario, a small simulation focusing on a sparse data setting is performed.
For simplicity, a binary treatment $A \sim \text{Bernoulli}(0.5)$ is considered. A binary covariate $X_1$ and a categorical covariate $X_2$ consisting of three levels are assumed and their distributions vary across sites: for sites 3, 6, and 9, $X_1 = 1, X_2 \sim  \text{Multinomial}(1; 0, 0.5, 0.5)$; for sites 1, 4, 7, and 10, $X_1 = 0, X_2 \sim \text{Multinomial}(1; 0.5, 0, 0.5) $; for sites 2, 5, and 8, $X_1 \sim \text{Bernoulli}(0.5), X_2 \sim \text{Multinomial}(1; 1/3, 1/3, 1/3)$.
The continuous outcome for an individual at site $i$ is generated by 
 $   Y_i = \beta_{i0} + \beta_{i1} x_1 + \beta_{i2} x_{2,2} + \beta_{i3} x_{2,3} + a(\psi_{i0} + \psi_{i1} x_1 + \psi_{i2} x_{2,2} + \psi_{i3} x_{2,3}) + \epsilon$,
where the random error $\epsilon$ follows a normal distribution with mean 0 and residual variance $\sigma_\epsilon^2=0.25$. For $X_2$, the first category is assumed as the reference level,  and two binary indicators $X_{2,2}$ and $X_{2,3}$ are created for the second and third categories of $X_2$. 
Two different settings are considered for the site-specific parameters $\boldsymbol{\theta_i} = (\beta_{i0}, \beta_{i1}, \beta_{i2}, \beta_{i3}, \psi_{i0}, \psi_{i1}, \psi_{i2}, \psi_{i3})$: 

\begin{itemize}
    \item[--] common effect: $\boldsymbol{\theta_1} = \boldsymbol{\theta_2} = \ldots = \boldsymbol{\theta_{10}}=\boldsymbol{\theta}$ and $\boldsymbol{\theta}=(\beta_0, \beta_1, \beta_2, \beta_3, \psi_0, \psi_1, \psi_2, \psi_3)$ is the common population parameter;
    \item[--] varying effects:  $ \boldsymbol{\theta_{i}} \sim MVN(\boldsymbol{\theta}, \boldsymbol{\Sigma_{\theta}})$, where $\boldsymbol{\Sigma_\theta}$ is a $8\times 8$ diagonal matrix where the between-site variance is obtained from the heterogeneity level $I^2 = \frac{\sigma_B^2}{\sigma_B^2 + \sigma_\epsilon^2} = 0.1, 0.2, 0.3$.
\end{itemize}

In both settings, the common treatment-free parameters are $\beta_0 = 4$, $\beta_1 =1 $, $\beta_2 =1$, $\beta_3=-1$,   and the common blip parameters are $\psi_0 = 1$, $\psi_1 =1$, $\psi_2 = -2.5$, $\psi_3=2$. The common optimal ITR is $d^{opt}(\boldsymbol{x}) = I(\psi_0 + \psi_1 x_1 + \psi_2 x_{2,2} + \psi_3 x_{2,3} > 0)$. The model details in a sparse data setting is described in Appendix \ref{appendix1}.

\subsubsection{Many covariates setting}
We consider two scenarios: a total of either 10 or 20 covariates is collected, but only three covariates,  $X_1$, $X_2$,  and $X_3$ are related to optimal treatment assignment. The covariates $X_1$ and $X_2$ are generated in the same way as in the binary treatment setting. We generate $X_3$ by the following distribution: for sites 3, 6, and 9, $X_3 \sim \text{Exponential}(1)$; for sites 1, 4, 7, and 10, $X_3 \sim \text{Exponential}(1.7)$; for sites 2, 5, and 7, $X_3 \sim \text{Exponential}(0.7)$. 
Other covariates are generated by $X_j \sim N(0,1)$, for $j\geq 4$.  For simplicity, we assume a binary treatment $A \sim \text{Bernoulli}(0.5)$.  The continuous outcome for an individual at site $i$ is generated by $Y_i = \beta_{i0} + \sum_{s=1}^p \beta_{is}x_{is} + A (\psi_{i0} + \sum_{t=1}^p \psi_{it}x_{it}) + \epsilon $, for $p=10$ or 20, and $\epsilon \sim N(0, 0.25)$.   The site-specific parameters $\beta_{is}$ and $\psi_{it}$ are generated under the assumption of varying effects: $\boldsymbol{\theta_i}=(\beta_{i0},\ldots, \beta_{ip}, \psi_{i0},\ldots, \psi_{ip}) \sim MVN (\boldsymbol{\theta}, \boldsymbol{\Sigma_\theta})$, where  $\boldsymbol{\theta}= (\beta_0,\ldots, \beta_p, \psi_0,\ldots,\psi_p)$ is a vector of common parameters and $\beta_0 = 4$, $\beta_{s}=1$ for $s\geq 2$, $\psi_{0} = 2.5, \psi_1 = -0.5, \psi_2 = 2, \psi_3 = -1$,  and $\psi_{t} = 0$  for $t\geq 4$.  The common optimal ITR is thus $d^{opt}(\boldsymbol{x}) = I(\psi_0 + \psi_1 x_1 + \psi_2 x_{2} + \psi_3 x_{2} > 0)$.  The between-site variance in the diagonal variance-covariance matrix $\boldsymbol{\Sigma_\theta}$  is obtained from the heterogeneity level $I^2 = \frac{\sigma_B^2}{\sigma_B^2 + \sigma_\epsilon^2} = 0.1$.

\subsection{Estimands, methods, and performance metrics} \label{subsec33}
The estimands of interest are the common blip parameters which fully characterize the optimal ITR. All analyses rely on the two-stage IPD meta-analysis, using linear regression as the stage-one model. For all scenarios, we use a Bayesian hierarchical model for the second stage. For the mean parameters in all settings other than the many covariates setting, we use a normal prior with mean 0 and variance 10,000. In the setting of many covariates, we assign the same normal prior to the common main treatment effect parameter but a horseshoe prior to all treatment-covariate interactions, selecting only those whose 95\% posterior credible intervals do not include zero. For variance component parameters, three priors with varying levels of informativeness are considered: 
half-Cauchy priors with location 0 and scale parameters 1, 10, or 100. However, in the many covariates setting, only a half-Cauchy (0,1) prior is used for the variance component parameters.

For all scenarios, 2000 iterations are performed. Measures of performance used to assess the ITR estimation are: (i) the relative bias of estimators of the blip parameters, which represents the difference between the mean of the estimates and the true value, divided by the latter, (ii) the standard deviation of the estimators, (iii) the difference between the value function (dVF) under the true optimal ITR and the value function under the estimated optimal ITR, 
 where the value function with respect to an ITR is the expected outcome if all patients in a population (in our simulation, it is a new cohort of patients of size $n=100,000$) were treated according to the ITR,
and (iv) the empirical standard deviation of the value function difference  when the estimated treatment rule was applied to the same population. For the many covariates setting, these measures are assessed over: (1) a full set of 2000 iterations, and (2) a subset of iterations where the non-zero treatment-covariate interactions are correctly  selected.  The proportion of selection, calculated as the number of times the covariate is selected divided by the total number of simulation iterations, is also measured.  The results are compared with results obtained from a one-stage approach based on the full individual-level data.

\section{Model details in a sparse data setting in simulation}
\label{appendix1}

To illustrate how the proposed two-stage approach can deal with data sparsity, a small simulation is performed and a binary treatment $A \in \{0,1\}$ is considered for simplicity. A binary covariate $X_1$ and a categorical covariate $X_2$ consisting of three levels are assumed: for the $i$-th site,
\[X_1 \sim \begin{cases} 
      1 & i = 3s \\
      0 & i = 3s + 1 \\
     \text{Bernoulli}(0.5) & i = 3s + 2\\
   \end{cases}, \quad 
   X_2 \sim \begin{cases} 
      \text{Multinomial}(1; 0, 0.5, 0.5) & i = 3s \\
      \text{Multinomial}(1; 0.5, 0, 0.5) & i = 3s + 1 \\
      \text{Multinomial}(1; 1/3, 1/3, 1/3) & i = 3s + 2\\
   \end{cases},
\]
where $s=0,\ldots,3$. The continuous outcome for an individual at site $i$ is generated by 
\begin{align}
    Y_i = \beta_{i0} + \beta_{i1} x_1 + \beta_{i2} x_{2,2} + \beta_{i3} x_{2,3} + a(\psi_{i0} + \psi_{i1} x_1 + \psi_{i2} x_{2,2} + \psi_{i3} x_{2,3}) + \epsilon,
    \label{eq12}
\end{align}
where the random error $\epsilon$ follows a normal distribution with mean 0 and residual variance $\sigma_\epsilon^2=0.25$. For $X_2$, the first category is assumed as the reference level,  and two binary indicators $X_{2,2}$ and $X_{2,3}$ are created for the second and third categories of $X_2$. As discussed in the main manuscript, both common effect and varying effects settings are explored for the site-specific parameters $\boldsymbol{\theta_i} = (\beta_{i0}, \beta_{i1}, \beta_{i2}, \beta_{i3}, \psi_{i0}, \psi_{i1}, \psi_{i2}, \psi_{i3})$.

Due to the data-generating mechanism, the implied (correctly-specified)  linear regression models at the first stage for sites $i=3s$, $s=1,2, 3$, are 
    \begin{align}
            E(Y_i\vert \boldsymbol{x},a) = \gamma_{i0} 
            +  \gamma_{i2}x_{2,2}
            + a(\xi_{i0} 
            + \xi_{i2}x_{2,2} 
            ).
            \label{surrogate1}
    \end{align}
    Since no patients within sites $i=3s$ are in the reference category of $X_2$, in site-specific analyses, one of $X_{2,2}$ and $X_{2,3}$ will be chosen as the new reference category, and its main effect as well as the interaction effect with the treatment in (\ref{eq12}) cannot be estimated. Here, without loss of generality, we assume that $X_{2,3}$ is the new reference category. Then, the site-specific main effect estimator of $X_{2,2}$, $\hat{\gamma}_{i2}$,  and its interaction effect estimator, $\hat{\xi}_{i2}$, in (\ref{surrogate1}) will be biased for $\beta_{i2}$ and $\psi_{i2}$ respectively, as $\gamma_{i2} = \beta_{i2} - \beta_{i3}$ and $\xi_{i2} = \psi_{i2} - \psi_{i3}$. In addition, since $X_1=1$ for all patients within sites $i=3s$, the effect of $X_1$ (i.e., $\beta_{i1}$) and its interaction with treatment (i.e., $\psi_{i1}$) cannot be estimated. The site-specific  intercept and main treatment effect estimators in (\ref{surrogate1}) (i.e., $\hat{\gamma}_{i0}$ and $\hat{\xi}_{i0}$) are biased for the original parameters $\beta_{i0}$ and $\psi_{i0}$ in (\ref{eq12}), as $\gamma_{i0}=\beta_{i0} + \beta_{i1} + \beta_{i3}$ and $\xi_{i0}=\psi_{i0} + \psi_{i1} + \psi_{i3}$.  Therefore, to recover the original parameters, the likelihood model for these sites at the second stage should be reparametrized as 
    \begin{align*}
        \hat{\xi}_{i0} &\sim N\left(\psi_{i0}+ \psi_{i1} + \psi_{i3}, sd(\hat{\xi}_{i0})^2 \right), &
        \hat{\xi}_{i2} & \sim N\left(\psi_{i2} -  \psi_{i3}, sd(\hat{\xi}_{i2})^2 
        \right),\\
        \psi_{i0} &\sim N\left(\psi_{0},\sigma_{\psi_0}^2\right), & \psi_{i1} &\sim N\left(\psi_{1}, \sigma_{\psi_1}^2\right), \\
        \psi_{i2} &\sim N\left(\psi_{2},\sigma_{\psi_2}^2\right),& \psi_{i3}  &\sim N\left(\psi_{3},\sigma_{\psi_3}^2\right).
    \end{align*}
 For sites $i=3s+1$, $s=0,\ldots,3$, the site-specific linear regression models are 
     \begin{align}
    E(Y_i \vert \boldsymbol{x},a) = \gamma_{i0} 
    + \gamma_{i3} x_{2,3} + a(\xi_{i0} 
    +  \xi_{i3} x_{2,3}).
    \label{surrogate2}
\end{align}
In these sites, $X_1=0$ for all patients, and no patients are in the second category of $X_2$. Therefore, $\beta_{i1},\beta_{i2},\psi_{i1},\psi_{i2}$ cannot be estimated. However, the estimators $\hat{\gamma}_{i0}$, $\hat{\gamma}_{i3}$, $\hat{\xi}_{i0}$ and $\hat{\xi}_{i3}$ in (\ref{surrogate2}) are still be consistent for the parameters $\beta_{i0}$, $\beta_{i3}$, $\psi_{i0}$ and $\psi_{i3}$ in (\ref{eq12}), as there are patients with $X_1=0$ and $X_{2,2}=X_{2,3}=0$. Thus, the likelihood model for these sites at the second stage will be 
\begin{align*}
         \hat{\xi}_{i0} &\sim N\left(\psi_{i0}, sd(\hat{\xi}_{i0})^2 
         \right), &
        \hat{\xi}_{i3} & \sim N\left(\psi_{i3} , sd(\hat{\xi}_{i3})^2 
        \right), \\
        \psi_{i0} &\sim N\left(\psi_{0},\sigma_{\psi_0}^2\right), & 
        \psi_{i3}&  \sim N\left(\psi_{3},\sigma_{\psi_3}^2\right).
\end{align*}

For sites $i=3s+2$, $s=0,1,2$, all levels of all covariates are represented and thus all parameters are estimable. The regression estimators in 
     \begin{align*}
    E(Y_i \vert \boldsymbol{x},a) = \gamma_{i0} + \gamma_{i1} x_1 + \gamma_{i2} x_{2,2} + \gamma_{i3} x_{2,3} + a(\xi_{i0} + \xi_{i1} x_1 + \xi_{i2} x_{2,2} +  \xi_{i3} x_{2,3})
\end{align*}
will be consistent for the corresponding parameters in (\ref{eq12}). The likelihood model at the second stage will be 
\begin{align*}
        \hat{\xi}_{i0} &\sim N\left(\psi_{i0}, sd(\hat{\xi}_{i0})^2  \right), &
        \hat{\xi}_{i1} &\sim N\left(\psi_{i1}, sd(\hat{\xi}_{i1})^2  \right),\\
        \hat{\xi}_{i2} &\sim N\left(\psi_{i2}, sd(\hat{\xi}_{i2})^2  \right), &
        \hat{\xi}_{i3} & \sim N\left(\psi_{i3} , sd(\hat{\xi}_{i3})^2  \right),\\
          \psi_{i0} &\sim N\left(\psi_{0},\sigma_{\psi_0}^2\right),& \psi_{i1} &\sim N\left(\psi_{1},\sigma_{\psi_1}^2\right), \\
        \psi_{i2} &\sim N\left(\psi_{2},\sigma_{\psi_2}^2\right),& \psi_{i3}  &\sim N\left(\psi_{3},\sigma_{\psi_3}^2\right).
\end{align*}

\section{Additional simulation results}
\label{appendix2}

\begin{sidewaysfigure}
    \centering
    \includegraphics[width=\textwidth]{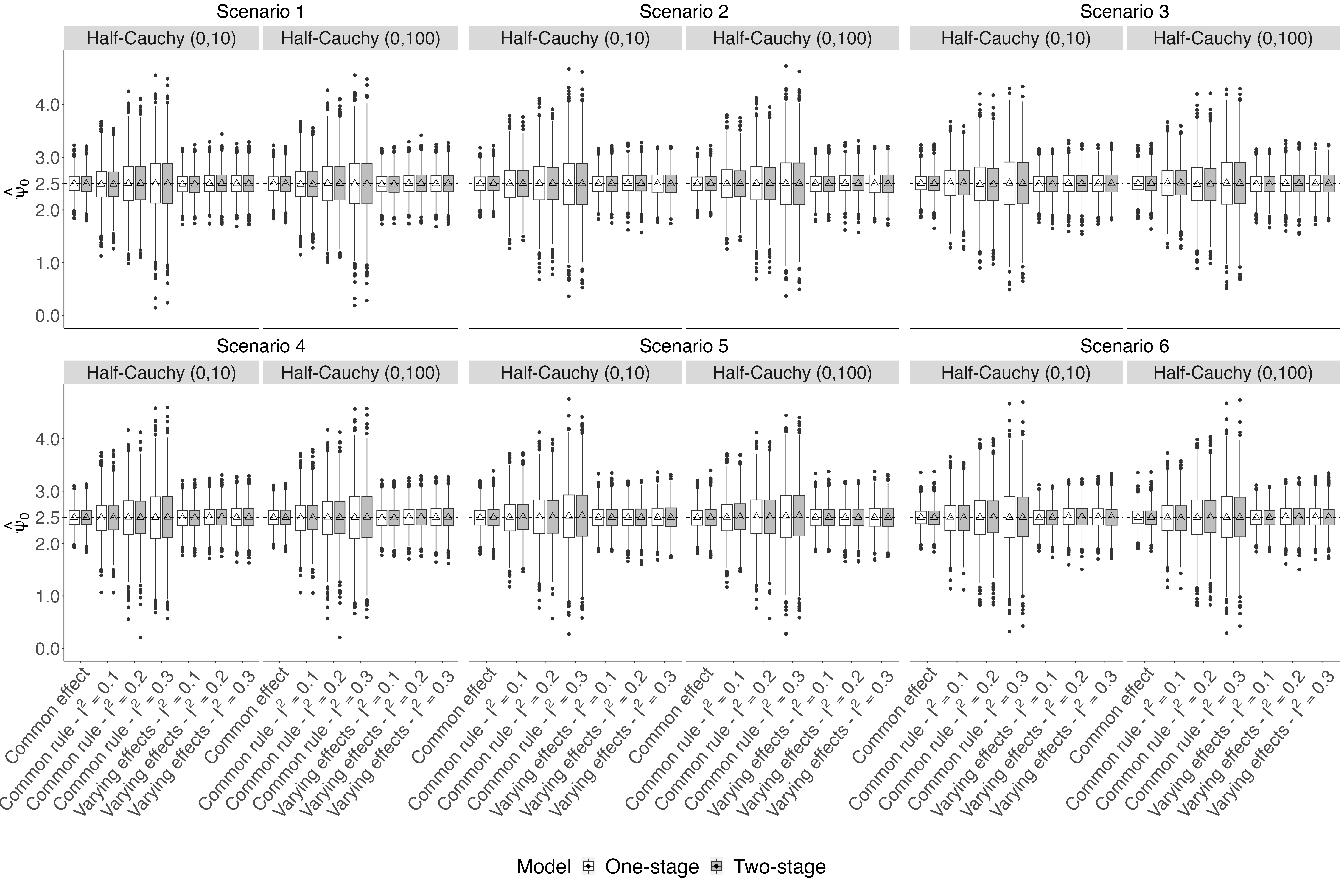}
    \caption{Simulation results for the small sample size and the binary treatment setting. Performance of the methods is assessed over 2000 iterations. Estimates (posterior means) of $\psi_0$ are shown under different confounding scenarios, heterogeneity levels ($I^2 = 0.1, 0.2, 0.3$),  and prior choices. The triangles represent the mean of the estimates in each case. The dashed line shows the true value of 2.5.}
    \label{fig:n=50-binary-psi0}
\end{sidewaysfigure}

\begin{sidewaysfigure}
    \centering
    \includegraphics[width=\textwidth]{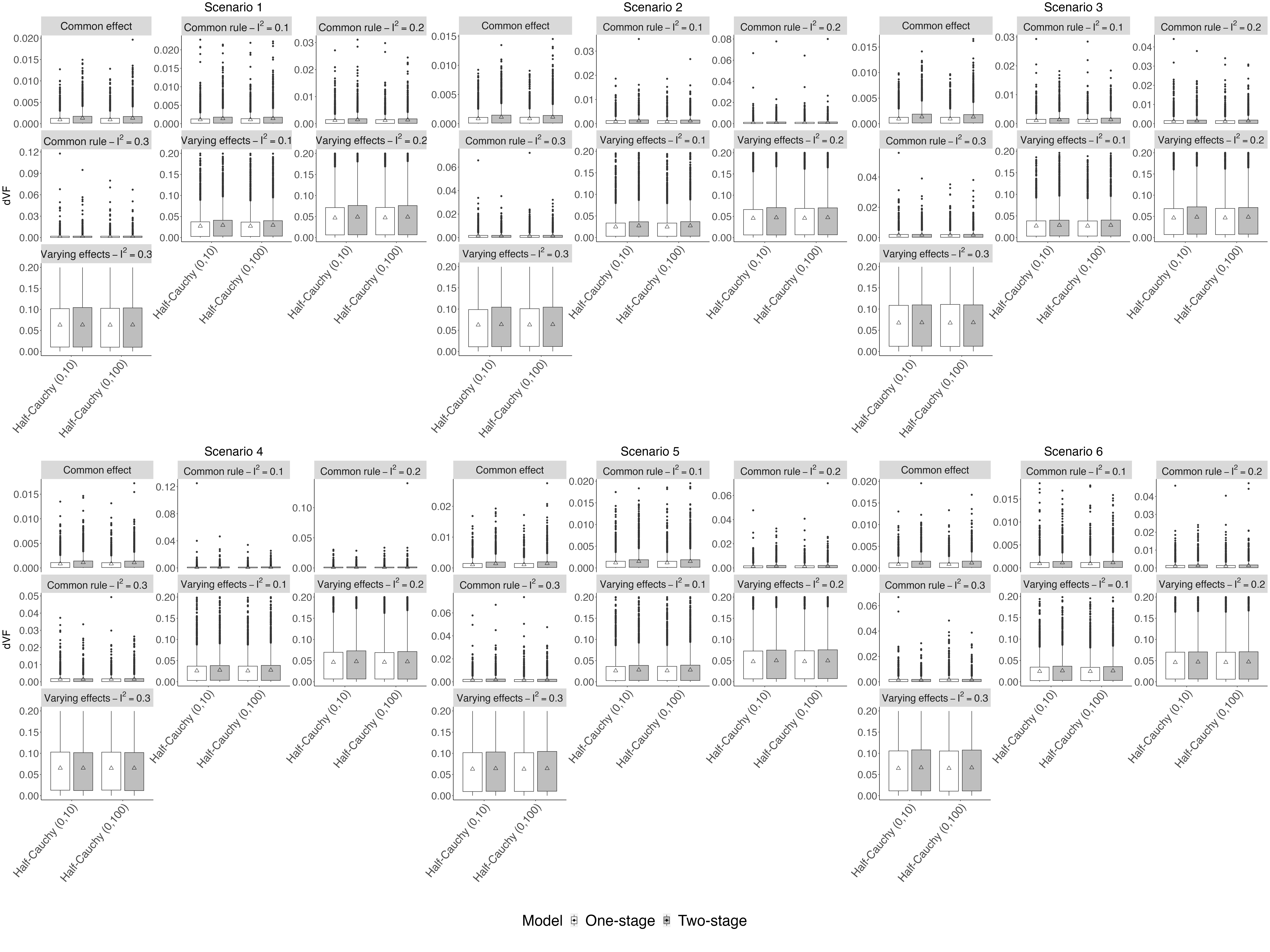}
    \caption{Simulation results for the small sample size and the binary treatment setting. Performance of the methods is assessed over 2000 iterations. The difference in the value function (dVF) between the true and estimated optimal ITR is shown under different confounding scenarios, heterogeneity levels ($I^2 = 0.1, 0.2, 0.3$),  and prior choices. The triangles represent the mean of the estimates in each case.}
    \label{fig:n=50-binary-dvf}
\end{sidewaysfigure}

\begin{sidewaysfigure}
    \centering
    \includegraphics[width=\textwidth]{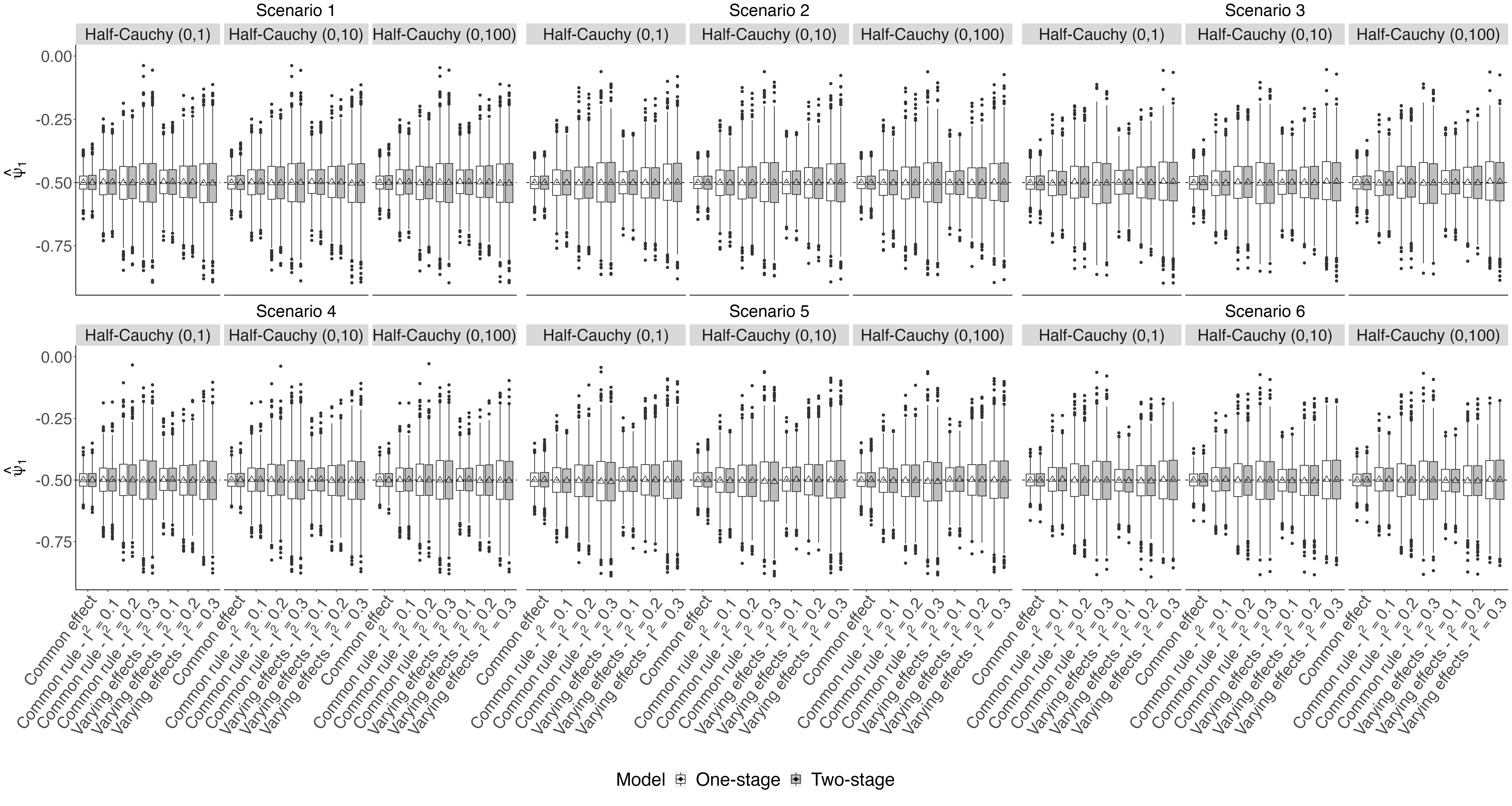}
    \caption{Simulation results for the small sample size and the binary treatment setting. Performance of the methods is assessed over 2000 iterations. Estimates (posterior means) of $\psi_1$ are shown under different confounding scenarios, heterogeneity levels ($I^2 = 0.1, 0.2, 0.3$),  and prior choices. The triangles represent the mean of the estimates in each case. The dashed line shows the true value of -0.5.}
    \label{fig:n=50-binary-psi1}
\end{sidewaysfigure}

\begin{sidewaysfigure}
    \centering
    \includegraphics[width=\textwidth]{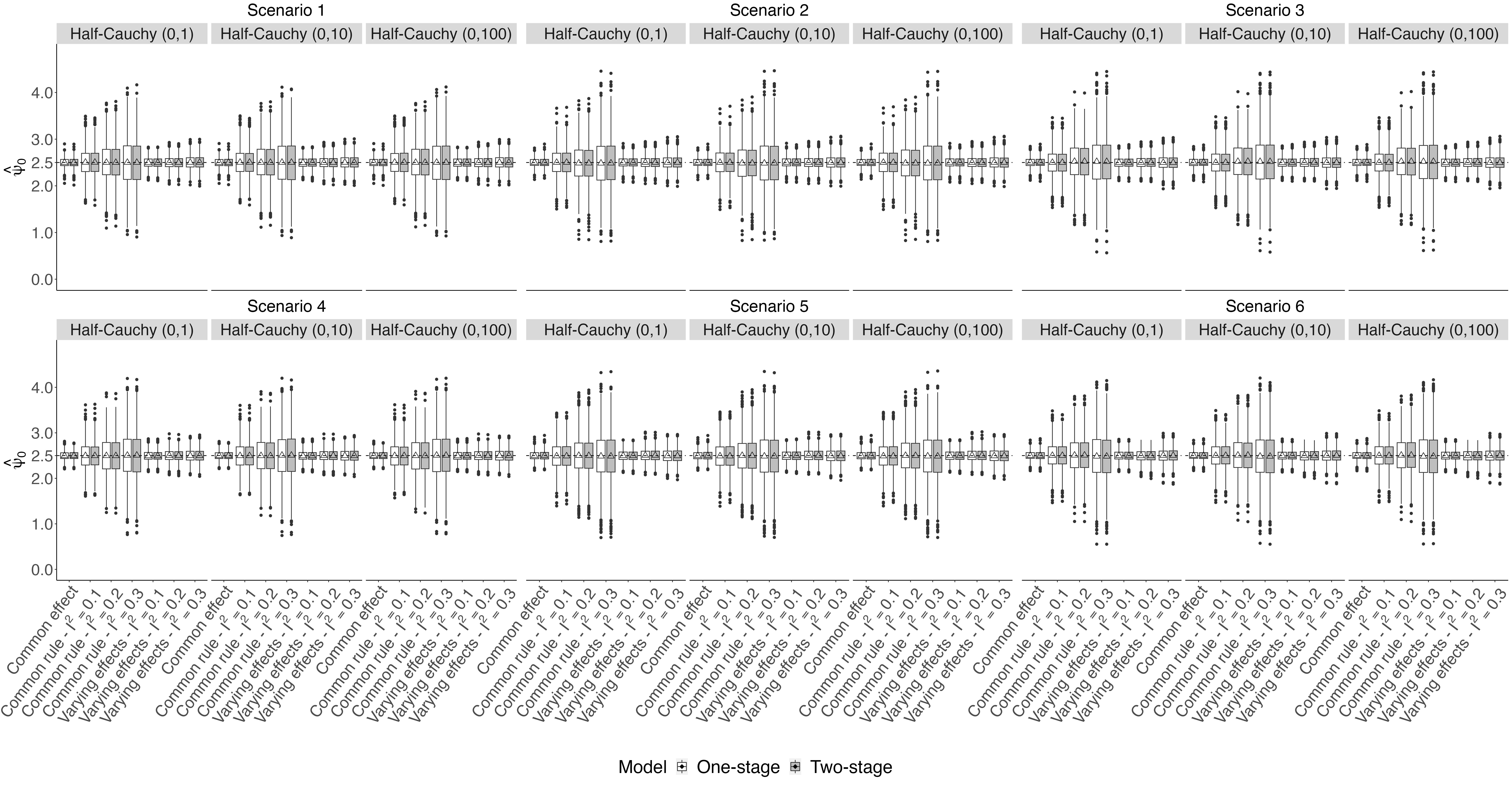}
    \caption{Simulation results for the large sample size and the binary treatment setting. Performance of the methods is assessed over 2000 iterations. Estimates (posterior means) of $\psi_0$ are shown under different confounding scenarios, heterogeneity levels ($I^2 = 0.1, 0.2, 0.3$),  and prior choices. The triangles represent the mean of the estimates in each case. The dashed line shows the true value of 2.5.}
    \label{fig:n=200-binary-psi0}
\end{sidewaysfigure}

\begin{sidewaysfigure}
    \centering
    \includegraphics[width=\textwidth]{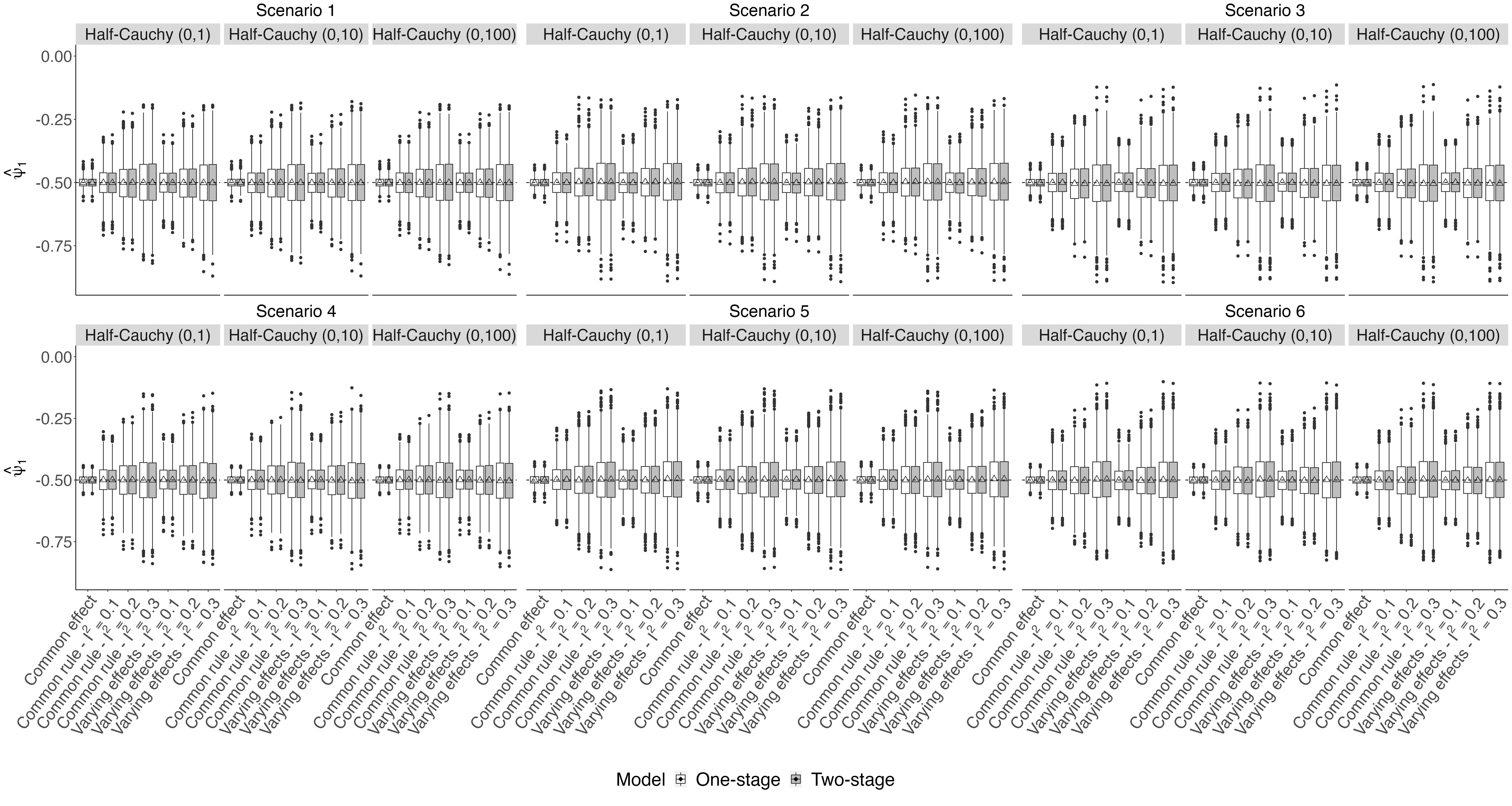}
    \caption{Simulation results for the large sample size and the binary treatment setting. Performance of the methods is assessed over 2000 iterations. Estimates (posterior means) of $\psi_1$ are shown under different confounding scenarios, heterogeneity levels ($I^2 = 0.1, 0.2, 0.3$),  and prior choices. The triangles represent the mean of the estimates in each case. The dashed line shows the true value of -0.5.}
    \label{fig:n=200-binary-psi1}
\end{sidewaysfigure}

\begin{sidewaysfigure}
    \centering
    \includegraphics[width=\textwidth]{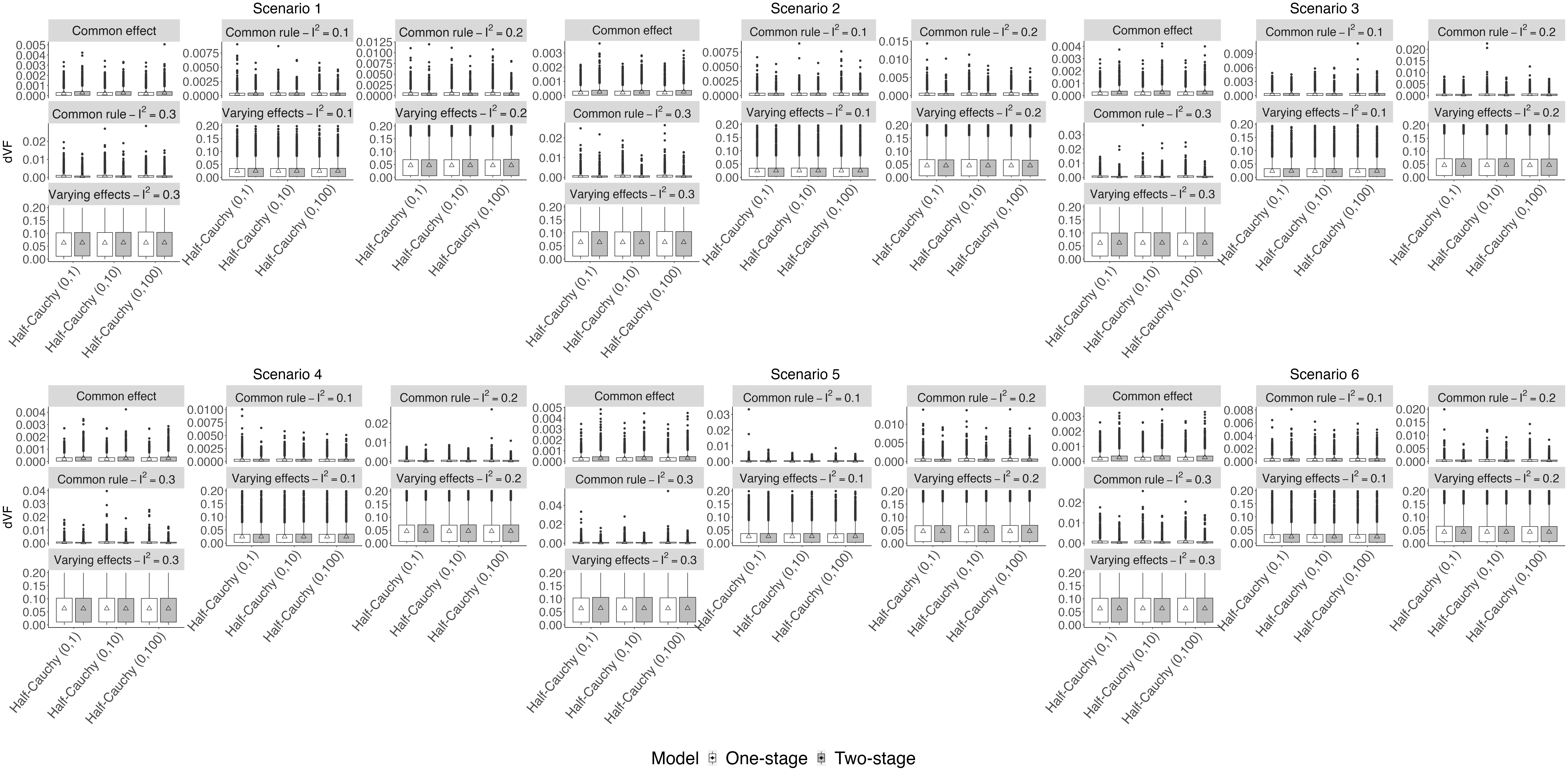}
    \caption{Simulation results for the large sample size and the binary treatment setting. Performance of the methods is assessed over 2000 iterations. The difference in the value function (dVF) between the true and estimated optimal ITR is shown under different confounding scenarios, heterogeneity levels ($I^2 = 0.1, 0.2, 0.3$),  and prior choices. The triangles represent the mean of the estimates in each case.}
    \label{fig:n=200-binary-dvf}
\end{sidewaysfigure}

\begin{figure}
    \centering
    \includegraphics[width=\textwidth]{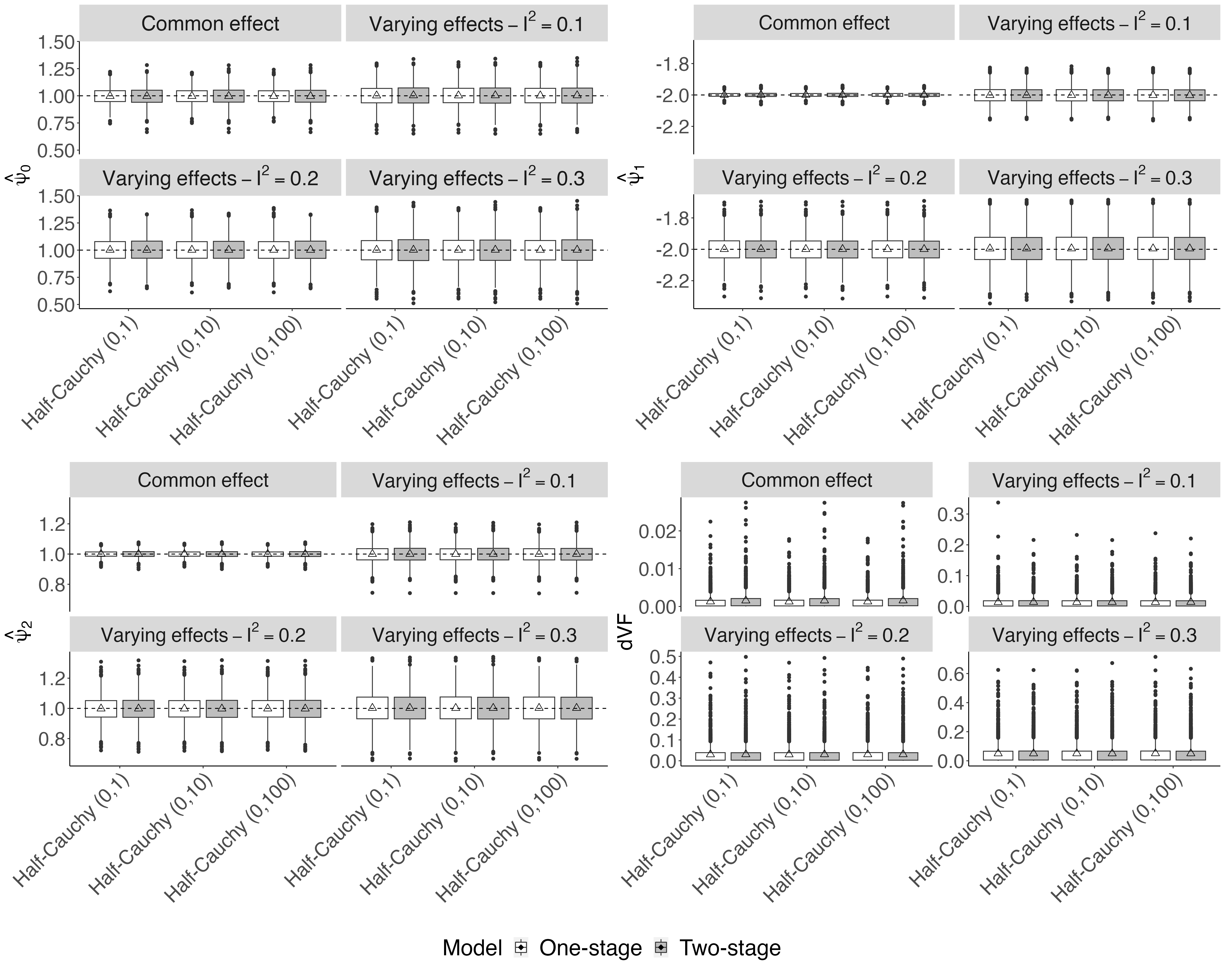}
    \caption{Simulation results for the small sample size and the continuous treatment setting. Performance of the methods is assessed over 2000 iterations. Estimates (posterior means) of $\psi_0$, $\psi_1$, $\psi_2$, and the difference in the value function (dVF) between the true and estimated optimal ITR are shown under different heterogeneity levels ($I^2 = 0.1, 0.2, 0.3$),  and prior choices. The triangles represent the mean of the estimates in each case. The dashed lines show the true values of $\psi_0$, $\psi_1$, $\psi_2$.}
    \label{fig:n=50-continuous}
\end{figure}

\begin{figure}
    \centering
    \includegraphics[width=\textwidth]{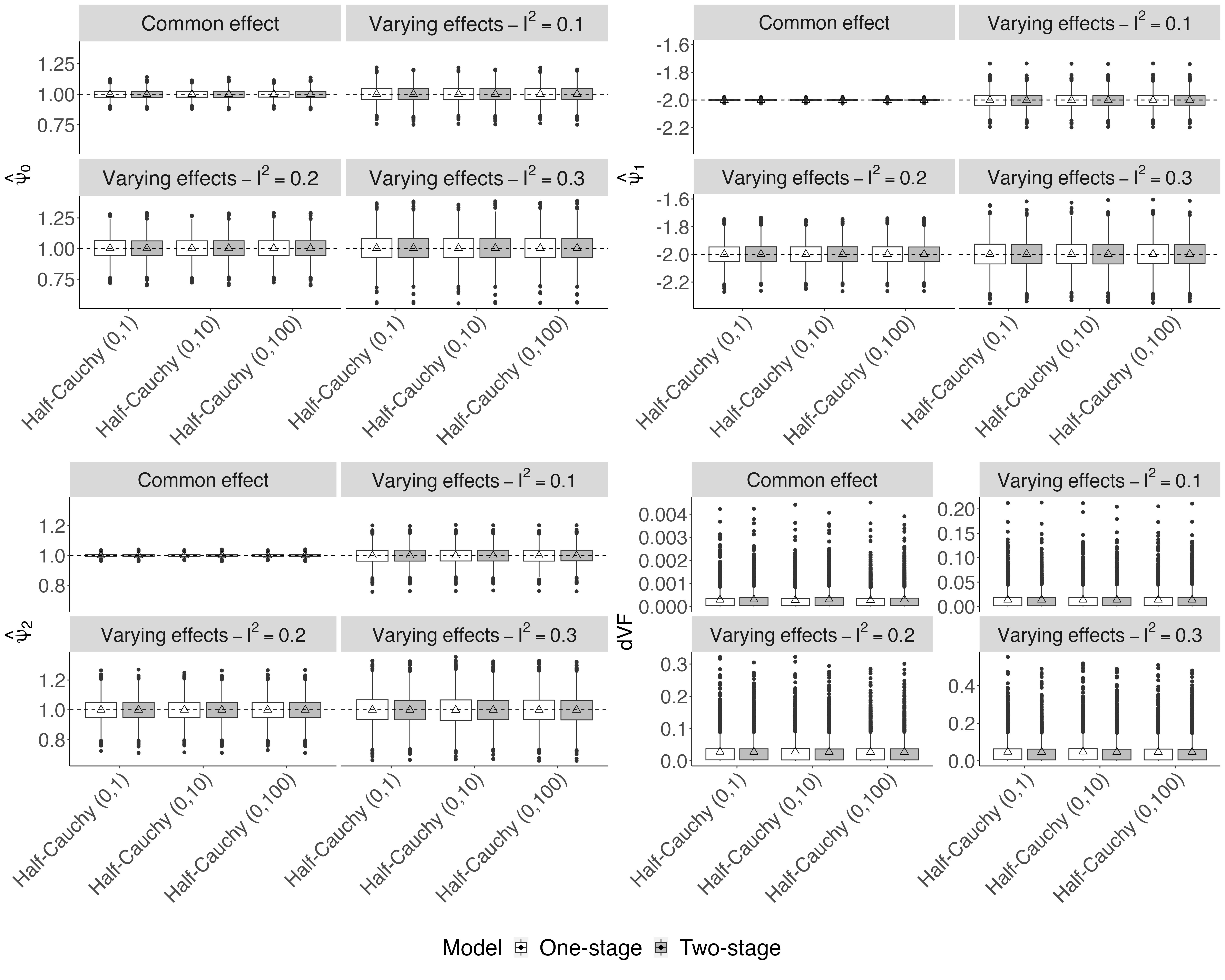}
    \caption{Simulation results for the large sample size and the continuous treatment setting. Performance of the methods is assessed over 2000 iterations. Estimates (posterior means) of $\psi_0$, $\psi_1$, $\psi_2$, and the difference in the value function (dVF) between the true and estimated optimal ITR are shown under different heterogeneity levels ($I^2 = 0.1, 0.2, 0.3$) and prior choices. The triangles represent the mean of the estimates in each case. The dashed lines show the true values of $\psi_0$, $\psi_1$, $\psi_2$. }
    \label{fig:n=200-continuous}
\end{figure}

\begin{figure}
    \centering
    \includegraphics[width=\textwidth]{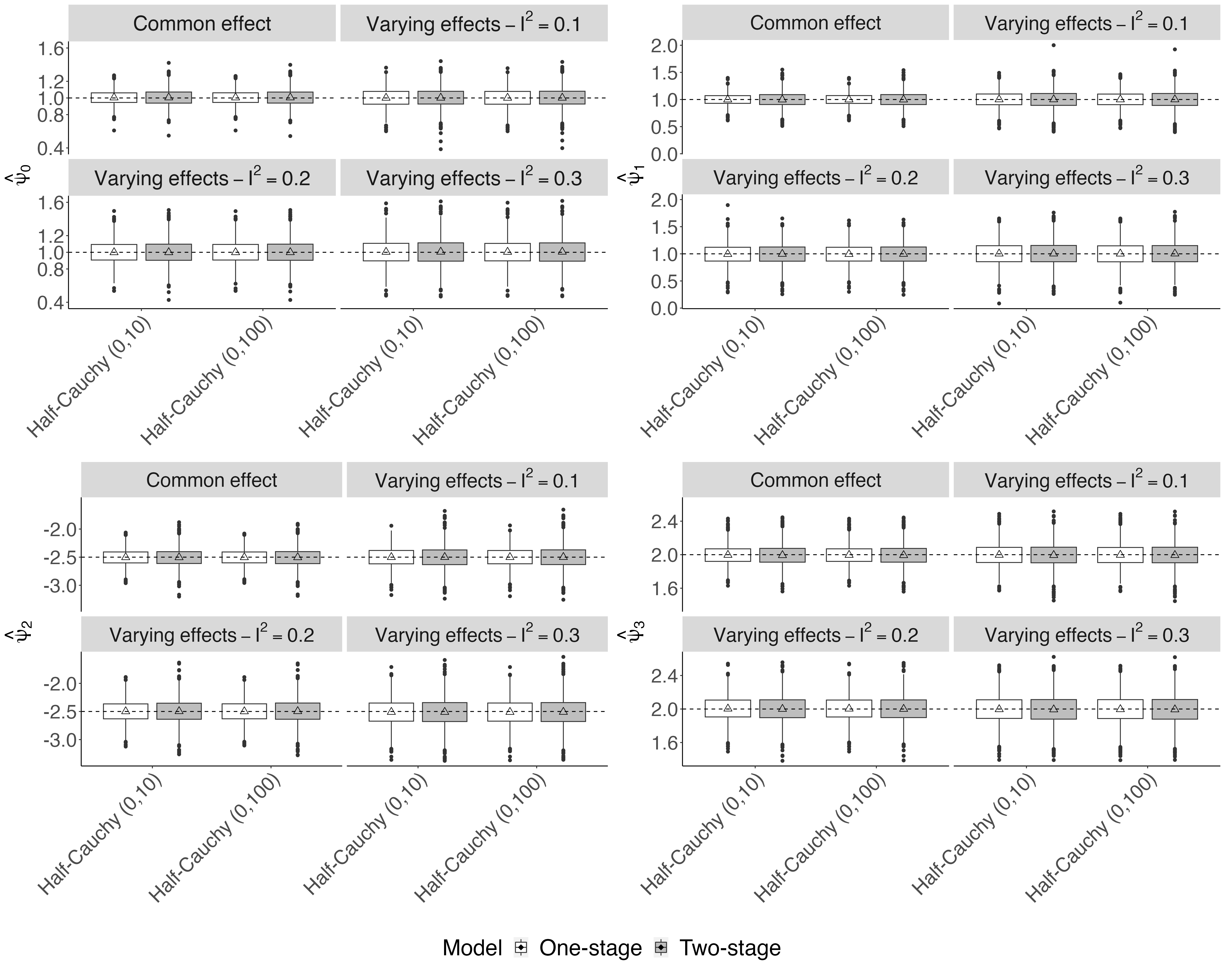}
    \caption{Simulation results for the small sample size and the sparse data setting. Performance of the methods is assessed over 2000 iterations. Estimates (posterior means) of $\psi_0$, $\psi_1$, $\psi_2$, $\psi_3$ are shown under different heterogeneity levels ($I^2 = 0.1, 0.2, 0.3$) and half-Cauchy (0,10) and half-Cauchy (0,100) priors. The triangles represent the mean of the estimates in each case. The dashed lines show the true values of $\psi_0$, $\psi_1$, $\psi_2$, $\psi_3$.}
    \label{fig:n=50-sparse}
\end{figure}

\begin{figure}
    \centering
    \includegraphics[width=\textwidth]{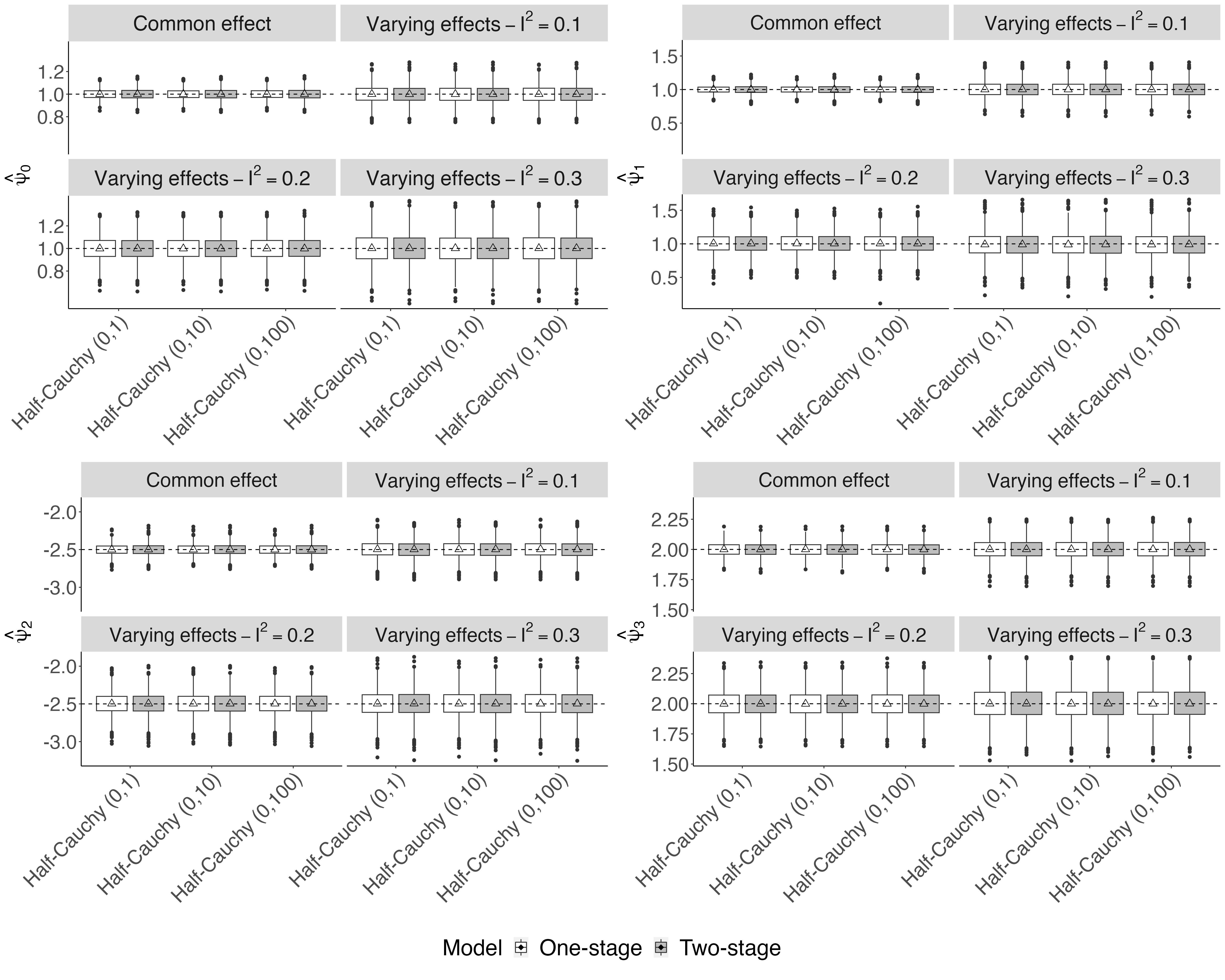}
    \caption{Simulation results for the large sample size and the sparse data setting. Performance of the methods is assessed over 2000 iterations. Estimates (posterior means) of $\psi_0$, $\psi_1$, $\psi_2$, $\psi_3$ are shown under different heterogeneity levels ($I^2 = 0.1, 0.2, 0.3$) and prior choices. The triangles represent the mean of the estimates in each case. The dashed lines show the true values of $\psi_0$, $\psi_1$, $\psi_2$, $\psi_3$.}
    \label{fig:n=200-sparse}
\end{figure}

This section presents additional results from the simulations carried out. Figures \ref{fig:n=50-binary-psi0} and \ref{fig:n=50-binary-dvf} shows estimates of $\psi_0$ and the dVF under half-Cauchy (0,10) and half-Cauchy (0,100) priors in the binary treatment setting with small sample size, and Figure \ref{fig:n=50-binary-psi1} shows simulation results for 
$\psi_1$. Figures \ref{fig:n=200-binary-psi0} - \ref{fig:n=200-binary-dvf} present simulation results for the large sample size and binary treatment setting, including the estimation of $(\psi_0,\psi_1)$ and the dVF. Figures \ref{fig:n=50-continuous} and \ref{fig:n=200-continuous} present the simulation results in the continuous treatment setting with both small and large sample sizes, including the estimation of $(\psi_0,\psi_1,\psi_2)$ and the dVF. Figures \ref{fig:n=50-sparse} and \ref{fig:n=200-sparse} present simulation results of $(\psi_0,\psi_1,\psi_2,\psi_3)$ in the sparse data setting with the large sample size. 

Similar patterns to those in the main manuscript are observed. The results are not sensitive to the different prior choices. The one- and two-stage approaches give similar results. Both provide unbiased estimations of blip function parameters. However, when the heterogeneity is small, the variability of estimators in the two-stage approach is larger than that in the one-stage approach, as the one stage approach is able to borrow information across sites when the heterogeneity is small. When the heterogeneity is large, this difference is smaller, and variability in both approaches increase compared to that with small heterogeneity.  In both binary and continuous treatment settings, the dVF increases with increasing heterogeneity, suggesting a worse optimal ITR estimation. In the sparse data setting, the dVF is zero or close to zero in all scenarios, regardless of the heterogeneity levels. We only consider binary covariates and binary treatment in the sparse data setting. The indicator function $I(\psi_0 + \psi_1 x_1 + \psi_2 x_{2,2} + \psi_3 x_{2,3} > 0)$ is less sensitive to the errors in the blip function parameter estimation compared with the optimal ITR in the setting of continuous treatment/covariates. Therefore, even if the parameter estimators are more varied with large heterogeneity, the dVF does not change much with different heterogeneity levels, and we (almost) obtain the true optimal ITR in all cases. Also, in all settings, with a larger sample size, we obtain a more precise optimal ITR estimation, as the variability of blip parameter estimation and the dVF are smaller.

\FloatBarrier
\section{Analysis of Warfarin data}
\label{appendix5}

\subsection{Visual inspection of the overlap assumption}
\begin{figure}[H]
    \centering
    \includegraphics[width=\textwidth]{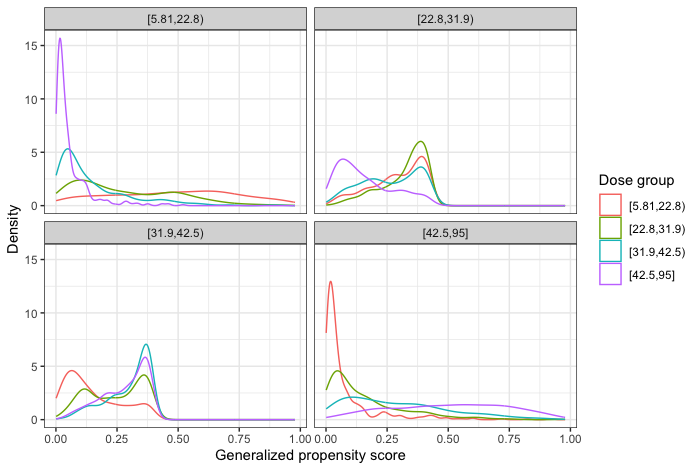}
    \caption{Distribution of the generalized propensity score by dose group in the Warfarin analysis.  }
    \label{fig:overlap}
\end{figure}

The overlap assessment was conducted by discretizing the continuous dose into four ordinal dose groups based on the minimum, 25-th, 50-th, and 75-th quantiles, and maximum of the observed doses.  Then, following that suggested by \cite{10.1214/19-AOAS1282},  a proportional odds logistic model including all potential confounders was used to estimate the generalized propensity score~\citep{imbens2000role}.    The distribution of the generalized propensity score is shown in Figure \ref{fig:overlap}. A moderate lack of overlap was observed, particularly within the dose groups $[5.81, 22.8)$ and $[42.5, 95]$.

\subsection{Details of the models in the Warfarin analysis}
As discussed in the main paper, the blip functions are quadratic in Warfarin dose. The linear regression model for site $i$ can be explicitly stated as:
\begin{align}
\begin{split}
   &  E(Y_i\vert \boldsymbol{x},a) = \beta_{i0} + \beta_{i1} \text{Age} + \beta_{i2} \text{Amiodarone} + \beta_{i3} \text{Female} \\
      &+ \beta_{i4} \text{Non-White} + \beta_{i5} \text{VKORC1(AG)} + \beta_{i6} \text{VKORC1(AA)} + \beta_{i7} \text{CYP2C9(12)}  \\
      & + \beta_{i8}\text{CYP2C9(other)} + \beta_{i9} \text{Weight} + \beta_{i10} \text{Height}\\
      &+ a \times \biggl\{  \psi_{i0}^{(1)} + \psi_{i1}^{(1)} \text{Age} + \psi_{i2}^{(1)} \text{Amiodarone} + \psi_{i3}^{(1)} \text{Female}  \\
      & + \psi_{i4}^{(1)} \text{Non-White}+ \psi_{i5}^{(1)} \text{VKORC1(AG)}+ \psi_{i6}^{(1)} \text{VKORC1(AA)} \\
      &+ \psi_{i7}^{(1)} \text{CYP2C9(12)} + \psi_{i8}^{(1)}\text{CYP2C9(other)} \biggl\}\\
      &+a^2 \times \biggl\{  \psi_{i0}^{(2)} + \psi_{i1}^{(2)} \text{Age} + \psi_{i2}^{(2)} \text{Amiodarone} + \psi_{i3}^{(2)} \text{Female}  \\
      & + \psi_{i4}^{(2)} \text{Non-White}+ \psi_{i5}^{(2)} \text{VKORC1(AG)}+ \psi_{i6}^{(2)} \text{VKORC1(AA)}\\
      &  + \psi_{i7}^{(2)} \text{CYP2C9(12)} + \psi_{i8}^{(2)}\text{CYP2C9(other)} \biggl\}.
\end{split}
\label{real-eq1}
\end{align}
We assume that $\psi_{it}^{(u)} \sim N(\psi_t^{(u)}, (\sigma_t^{(u)})^2)$, $t=0,\ldots,8$, $u=1,2$. The parameters of interests are the common blip function parameters $\boldsymbol{\psi^{(1)}}=(\psi_{0}^{(1)}, \ldots, \psi_{8}^{(1)})$ and $\boldsymbol{\psi^{(2)}}=(\psi_{0}^{(2)}, \ldots, \psi_{8}^{(2)})$ which fully characterize the optimal Warfarin dosing. The unknown between-site variability associated with $\psi_{it}^{(u)}$ is denoted by $(\sigma_t^{(u)})^2$.

Table \ref{warfarin-sitespecific} shows site-specific blip function parameter estimates obtained from the stage-one (frequentist) linear regression models and the associated standard deviations for the Warfarin data. Due to data sparsity, some site-specific blip function parameters cannot be estimated in some sites. As demonstrated in the simulation, we need to modify the proposed two-stage model.

For $i=2$, the linear regression model in the first stage is 
\begin{align}
    \begin{split}
         E(Y_i\vert \boldsymbol{x},a) &= \gamma_{i0} + \gamma_{i1} \text{Age} + \gamma_{i2} \text{Amiodarone} + \gamma_{i3} \text{Female}  + \gamma_{i5} \text{VKORC1(AG)} \\
      &+ \gamma_{i6} \text{VKORC1(AA)}  + \beta_{i8}\text{CYP2C9(other)} + \gamma_{i9} \text{Weight} + \gamma_{i10} \text{Height} \\
      &+ a \times \biggl\{  \xi_{i0}^{(1)} + \xi_{i1}^{(1)} \text{Age} + \xi_{i2}^{(1)} \text{Amiodarone} + \xi_{i3}^{(1)} \text{Female}  \\
      & + \xi_{i5}^{(1)} \text{VKORC1(AG)}+ \xi_{i6}^{(1)} \text{VKORC1(AA)}  + \xi_{i8}^{(1)}\text{CYP2C9(other)} \biggl\}\\
      &+a^2 \times \biggl\{  \xi_{i0}^{(2)} + \xi_{i1}^{(2)} \text{Age} + \xi_{i2}^{(2)} \text{Amiodarone} + \xi_{i3}^{(2)} \text{Female}  \\
      & + \xi_{i5}^{(2)} \text{VKORC1(AG)}+ \xi_{i6}^{(2)} \text{VKORC1(AA)} +  \xi_{i8}^{(2)}\text{CYP2C9(other)} \biggl\}.
    \end{split}
    \label{real-eq2}
\end{align}

The parameters in equation (\ref{real-eq2}) satisfy
\begin{align*}
    \gamma_{i0} &= \beta_{i0} + \beta_{i4}, & \xi_{i0}^{(1)} &= \psi_{i0}^{(1)} + \psi_{i4}^{(1)}, & \xi_{i0}^{(2)} &= \psi_{i0}^{(2)} + \psi_{i4}^{(2)},\\
      \gamma_{it} & = \beta_{it},  & \xi_{it}^{(1)} &= \psi_{it}^{(1)}, & \xi_{it}^{(2)} &= \psi_{it}^{(2)},
\end{align*}
for $t \neq 0,4,7,$ since all patients in Site 2 are non-White and none carry CYP2C9 genotype 12. The modified likelihood model in the second stage is then 
\begin{align*}
    \hat{\xi}_{i0}^{(u)} &\sim N\left(\psi_{i0}^{(u)} + \psi_{i4}^{(u)}, sd(\hat{\xi}_{i0}^{(u)})^2\right),\\
    \hat{\xi}_{it}^{(u)} & \sim N\left(\psi_{it}^{(u)}, sd(\hat{\xi}_{i0}^{(u)})^2\right),\\
    \psi_{il}^{(u)} &\sim N\left(\psi_l^{(u)}, (\sigma_{l}^{(u)})^2 \right),
\end{align*}
for $t \neq 0,4,7$, $l \neq 7$ and $u=1,2$.

For $i=4$, the linear regression model in the first stage is 
\begin{align}
    \begin{split}
         &E(Y_i \vert \boldsymbol{x},a) = \gamma_{i0} + \gamma_{i1} \text{Age} + \gamma_{i2} \text{Amiodarone} + \gamma_{i3} \text{Female}  + \gamma_{i5} \text{VKORC1(AG)} \\
      &  + \beta_{i8}\text{CYP2C9(other)} + \gamma_{i9} \text{Weight} + \gamma_{i10} \text{Height} \\
      &+ a \times \biggl\{  \xi_{i0}^{(1)} + \xi_{i1}^{(1)} \text{Age} + \xi_{i2}^{(1)} \text{Amiodarone} + \xi_{i3}^{(1)} \text{Female}  + \xi_{i5}^{(1)} \text{VKORC1(AG)} \\
      &  + \xi_{i8}^{(1)}\text{CYP2C9(other)} \biggl\}\\
      &+a^2 \times \biggl\{  \xi_{i0}^{(2)} + \xi_{i1}^{(2)} \text{Age} + \xi_{i2}^{(2)} \text{Amiodarone} + \xi_{i3}^{(2)} \text{Female}  + \xi_{i5}^{(2)} \text{VKORC1(AG)} \\
      & +  \xi_{i8}^{(2)}\text{CYP2C9(other)} \biggl\}.
    \end{split}
    \label{real-eq3}
\end{align}
The parameters in equation (\ref{real-eq3}) satisfy 
\begin{align*}
      \gamma_{i0} &= \beta_{i0} + \beta_{i4} + \beta_{i6}, & \xi_{i0}^{(1)} &= \psi_{i0}^{(1)} + \psi_{i4}^{(1)}+ \psi_{i6}^{(1)}, & \xi_{i0}^{(2)} &= \psi_{i0}^{(2)} + \psi_{i4}^{(2)} + \psi_{i6}^{(2)},\\
      \gamma_{i5} &= \beta_{i5} - \beta_{i6} , & \xi_{i5}^{(1)} &= \psi_{i5}^{(1)}  - \psi_{i6}^{(1)}, & \xi_{i5}^{(2)} &= \psi_{i5}^{(2)} - \psi_{i6}^{(2)},\\
      \gamma_{it} & = \beta_{it},  & \xi_{it}^{(1)} &= \psi_{it}^{(1)}, & \xi_{it}^{(2)} &= \psi_{it}^{(2)},
\end{align*}
for $t=1,2,3,8,$ since all patients in Site 4 are non-White, and none carry VKORC1 genotype GG or CYP2C9 genotype 12. The modified likelihood model is
\begin{align*}
        \hat{\xi}_{i0}^{(u)} &\sim N\left(\psi_{i0}^{(u)} + \psi_{i4}^{(u)} + \psi_{i6}^{(u)}, sd(\hat{\xi}_{i0}^{(u)})^2\right),\\
         \hat{\xi}_{i5}^{(u)} &\sim N\left(\psi_{i5}^{(u)} -\psi_{i6}^{(u)}, sd(\hat{\xi}_{i0}^{(u)})^2\right),\\
    \hat{\xi}_{it}^{(u)} & \sim N\left(\psi_{it}^{(u)}, sd(\hat{\xi}_{i0}^{(u)})^2\right),\\
    \psi_{il}^{(u)} &\sim N\left(\psi_l^{(u)}, (\sigma_{l}^{(u)})^2 \right),
\end{align*}
for $t=1,2,3,8$, $l\neq 7$, $u=1,2$.

For $i=5$, the linear regression model in the first stage is 
\begin{align}
    \begin{split}
         E(Y_i\vert \boldsymbol{x},a) &= \gamma_{i0} + \gamma_{i1} \text{Age}  + \gamma_{i3} \text{Female}  + \gamma_{i5} \text{VKORC1(AG)} \\
      & + \gamma_{i7} \text{CYP2C9(12)} + \beta_{i8}\text{CYP2C9(other)} + \gamma_{i9} \text{Weight} + \gamma_{i10} \text{Height} \\
      &+ a \times \biggl\{  \xi_{i0}^{(1)} + \xi_{i1}^{(1)} \text{Age}  + \xi_{i3}^{(1)} \text{Female}  + \xi_{i5}^{(1)} \text{VKORC1(AG)} \\
      &  +\xi_{i7}^{(1)}\text{CYP2C9(12)} + \xi_{i8}^{(1)}\text{CYP2C9(other)} \biggl\}\\
      &+a^2 \times \biggl\{  \xi_{i0}^{(2)} + \xi_{i1}^{(2)} \text{Age}  + \xi_{i3}^{(2)} \text{Female}  + \xi_{i5}^{(2)} \text{VKORC1(AG)} \\
      & +\xi_{i7}^{(2)}\text{CYP2C9(12)} +  \xi_{i8}^{(2)}\text{CYP2C9(other)} \biggl\}.
    \end{split}
    \label{real-eq4}
\end{align}
The parameters in equation (\ref{real-eq4}) satisfy
\begin{align*}
       \gamma_{i0} &= \beta_{i0} + \beta_{i4}, & \xi_{i0}^{(1)} &= \psi_{i0}^{(1)} + \psi_{i4}^{(1)}, & \xi_{i0}^{(2)} &= \psi_{i0}^{(2)} + \psi_{i4}^{(2)} ,\\
      \gamma_{it} & = \beta_{it},  & \xi_{it}^{(1)} &= \psi_{it}^{(1)}, & \xi_{it}^{(2)} &= \psi_{it}^{(2)},
\end{align*}
for $t \neq 0,2, 4,6,$ since all patients in Site 5 are non-White and none take amiodarone or carry VKORC1 genotype AA. The modified likelihood model is  
\begin{align*}
        \hat{\xi}_{i0}^{(u)} &\sim N\left(\psi_{i0}^{(u)} + \psi_{i4}^{(u)} , sd(\hat{\xi}_{i0}^{(u)})^2\right),\\
    \hat{\xi}_{it}^{(u)} & \sim N\left(\psi_{it}^{(u)}, sd(\hat{\xi}_{i0}^{(u)})^2\right),\\
    \psi_{il}^{(u)} &\sim N\left(\psi_l^{(u)}, (\sigma_{l}^{(u)})^2 \right),
\end{align*}
for $t \neq 0, 2, 4,6$, $l \neq 2,6$, $u=1,2.$

For $i=6,8$, the linear regression model in the first stage is 
\begin{align}
    \begin{split}
         E(Y_i\vert \boldsymbol{x},a) &= \gamma_{i0} + \gamma_{i1} \text{Age} + \gamma_{i2} \text{Amiodarone} + \gamma_{i3} \text{Female}  + \gamma_{i5} \text{VKORC1(AG)} \\
      &+ \gamma_{i6} \text{VKORC1(AA)}  + \gamma_{i7} \text{CYP2C9(12)} + \beta_{i8}\text{CYP2C9(other)} + \gamma_{i9} \text{Weight} + \gamma_{i10} \text{Height} \\
      &+ a \times \biggl\{  \xi_{i0}^{(1)} + \xi_{i1}^{(1)} \text{Age} + \xi_{i2}^{(1)} \text{Amiodarone} + \xi_{i3}^{(1)} \text{Female}  + \xi_{i5}^{(1)} \text{VKORC1(AG)} \\
      &+ \xi_{i6}^{(1)} \text{VKORC1(AA)}  + \xi_{i7}^{(1)} \text{CYP2C9(12)} + \xi_{i8}^{(1)}\text{CYP2C9(other)} \biggl\}\\
      &+a^2 \times \biggl\{  \xi_{i0}^{(2)} + \xi_{i1}^{(2)} \text{Age} + \xi_{i2}^{(2)} \text{Amiodarone} + \xi_{i3}^{(2)} \text{Female}  + \xi_{i5}^{(2)} \text{VKORC1(AG)} \\
      &+ \xi_{i6}^{(2)} \text{VKORC1(AA)} +  \xi_{i7}^{(2)} \text{CYP2C9(12)} + \xi_{i8}^{(2)}\text{CYP2C9(other)} \biggl\}.
    \end{split}
    \label{real-eq5}
\end{align}
The parameters in equation (\ref{real-eq5}) satisfy
\begin{align*}
      \gamma_{it} & = \beta_{it},  & \xi_{it}^{(1)} &= \psi_{it}^{(1)}, & \xi_{it}^{(2)} &= \psi_{it}^{(2)},
\end{align*}
for $t\neq 4$, since all patients in Sites 6 and 8 are White. The likelihood model is then
\begin{align*}
    \hat{\xi}_{it}^{(u)} & \sim N\left(\psi_{it}^{(u)}, sd(\hat{\xi}_{i0}^{(u)})^2\right),\\
    \psi_{it}^{(u)} &\sim N\left(\psi_t^{(u)}, (\sigma_{t}^{(u)})^2 \right),
\end{align*}
for $t\neq 4$, $u=1,2$.

For $i=1,3,7,9$, no modification is needed, as all levels of all covariates are represented in the sites' data. Therefore, the likelihood model is 
\begin{align*}
    \hat{\xi}_{it}^{(u)} & \sim N\left(\psi_{it}^{(u)}, sd(\hat{\xi}_{i0}^{(u)})^2\right),\\
    \psi_{it}^{(u)} &\sim N\left(\psi_t^{(u)}, (\sigma_{t}^{(u)})^2 \right),
\end{align*}
for $t=0,\ldots,8$, $u=1,2$.



\begin{table}[ht]
\centering
\caption{Site-specific blip function parameter estimates in the stage-one (frequentist) linear regression models and the associated standard deviations (in parantheses) in the analysis of International Warfarin Pharmacogenetics Consortium data. The results are rescaled by a factor of 1000.}
\label{warfarin-sitespecific}
\resizebox{\textwidth}{!}{
\begin{tabular}{lrrrrr}
\toprule
  $\boldsymbol{\hat{\xi}_i^{(1)}}$\\
  \toprule
 Site & 1 & 2 & 3 & 4 & 5   \\ 
  \midrule
Intercept & -51.19 (149.03) & 27.14 (45.78) & 9.60 (18.46) & 1.96 (41.61) & 4.01 (24.56)  \\ 
  Age & 3.26 (13.88) & -1.43 (6.57) & -1.18 (3.37) & 6.33 (9.76) & -2.46 (5.10)  \\ 
  Amiodarone & 20.10 (95.06) & -102.99 (82.50) & 11.19 (21.90) & -82.00 (155.67) &  NA \\ 
  Female & 43.38 (125.63) & -23.85 (17.18) & -5.31 (12.68) & -19.48 (19.06) & -7.57 (16.65)  \\ 
  Non-White & -2549.54 (18299.82) &  NA & 28.38 (20.21) &  NA  & NA   \\ 
  VKORC1 (AG) & 10.43 (112.64) & -3.60 (38.47) & -2.05 (7.78) & -10.33 (31.52) & -3.36 (22.95)  \\ 
  VKORC1 (AA) & -277.89 (642.82) & 22.33 (37.04) & 6.07 (25.17) & NA  &  NA   \\ 
  CYP2C9 (12) & 6.66 (48.22) &  NA & 9.55 (13.41) &  NA  & -61.24 (116.28)  \\ 
  CYP2C9 (other) & -167.04 (731.37) & 17.63 (37.39) & 11.83 (12.83) & -3.04 (53.03) & 75.87 (43.38) \\
  \midrule
  Site & 6 & 7 & 8 & 9\\
  \midrule
  Intercept &  -9.02 (32.16) & 6.86 (17.10) & 14.23 (12.31) & -6.83 (15.87) & \\ 
  Age & 1.16 (4.81) & -5.01 (2.87) & -4.10 (1.74) & 2.07 (2.82) & \\ 
  Amiodarone &  15.18 (24.78) & -16.65 (20.54) & -17.27 (19.31) & 6.00 (21.70) & \\ 
  Female & -4.54 (14.14) & 15.01 (8.08) & 5.70 (6.83) & -0.36 (7.97) &  \\ 
  Non-White &  NA  & -3.49 (7.69) &  NA  & 0.04 (10.61) &  \\ 
  VKORC1 (AG) & -3.27 (16.99) & 12.10 (7.54) & 3.48 (9.28) & 3.41 (10.20) &  \\ 
  VKORC1 (AA) &  -28.17 (76.13) & -13.08 (23.31) & -7.08 (15.56) & -58.08 (43.37) &  \\ 
  CYP2C9 (12) &  31.57 (14.58) & 12.84 (12.43) & -1.79 (8.46) & -10.57 (13.70) &  \\ 
  CYP2C9 (other) &  -15.36 (29.83) & -13.79 (9.43) & 4.83 (9.23) & 9.03 (21.34) & \\
   \bottomrule
   $\boldsymbol{\hat{\xi}_i^{(2)}}$\\
   \bottomrule
   Site & 1 & 2& 3&4 & 5\\
   \midrule
   Intercept & 0.54 (1.69) & -0.45 (0.67) & -0.04 (0.22) & 0.04 (0.66) & -0.06 (0.24)  \\ 
  Age & -0.02 (0.21) & 0.06 (0.12) & -0.01 (0.04) & -0.13 (0.16) & 0.03 (0.05)  \\ 
  Amiodarone & -0.29 (1.46) & 2.75 (2.45) & -0.13 (0.29) & 1.18 (2.71) &  NA   \\ 
  Female & -0.59 (1.78) & 0.50 (0.36) & 0.08 (0.16) & 0.32 (0.30) & 0.07 (0.16) \\ 
  Non-White & 30.81 (229.31) &  NA  & -0.24 (0.21) &  NA  &  NA  \\ 
  VKORC1 (AG) & 0.09 (1.47) & -0.15 (0.48) & -0.01 (0.09) & 0.34 (0.44) & 0.12 (0.28)  \\ 
  VKORC1 (AA) & 7.30 (15.69) & -0.82 (0.54) & -0.05 (0.49) & NA  &  NA  \\ 
  CYP2C9 (12) & -0.15 (0.54) &  NA & -0.12 (0.19) &  NA  & 1.22 (1.69)  \\ 
  CYP2C9 (other) & 4.58 (18.94) & -0.03 (0.89) & -0.21 (0.19) & 0.07 (1.15) & -0.96 (0.55)  \\ 
  \midrule
   Site & 6 & 7& 8&9 & \\
   \midrule
   Intercept & 0.07 (0.36) & -0.06 (0.21) & -0.22 (0.13) & 0.09 (0.16) \\ 
  Age & -0.01 (0.05) & 0.06 (0.04) & 0.05 (0.02) & -0.03 (0.03) \\ 
  Amiodarone &  -0.01 (0.27) & 0.19 (0.30) & 0.24 (0.29) & -0.09 (0.31) \\ 
  Female &  0.05 (0.18) & -0.21 (0.10) & -0.08 (0.07) & 0.02 (0.09) \\ 
  Non-White &   NA  & 0.03 (0.08) &  NA  & -0.03 (0.12) \\ 
  VKORC1 (AG) &  0.05 (0.19) & -0.15 (0.08) & 0.02 (0.10) & -0.06 (0.13) \\ 
  VKORC1 (AA) & 0.86 (1.70) & 0.20 (0.41) & 0.11 (0.23) & 1.22 (0.96) \\ 
  CYP2C9 (12) &  -0.42 (0.17) & -0.16 (0.14) & 0.01 (0.09) & 0.12 (0.18) \\ 
  CYP2C9 (other) &  0.26 (0.53) & 0.15 (0.10) & -0.00 (0.11) & -0.15 (0.34) \\ 
  \bottomrule
\end{tabular}}
\end{table}

To select the variables that are truly relevant for the treatment decision, a horseshoe prior~\citep{shrinkage} is assumed for all treatment-covariate interactions. Specifically, for $t=1,\ldots, 8$ and $u=1,2$, we have 
\begin{align*}
\begin{split}
     \psi_t^{(u)} & \sim N(0, \tau^2 (\lambda_t^{(u)})^2),\\
     \lambda_t^{(u)} &\sim \text{Half-Cauchy (0,1)},\\
     \tau & \sim \text{Half-Cauchy (0,1)},
     \end{split}
\end{align*}
where $\tau$ and $\lambda_t^{(u)}$ are, respectively, the global and local shrinkage parameters. If the 95\% credibility interval of $\psi_{t}^{(u)}$, $t=1,\ldots,8$, $u=1,2,$ contains zero, the corresponding treatment-covariate interaction will not be selected, suggesting that the associated covariate has no evidence of a  tailoring effect on the optimal Warfarin dosing. For $\psi_0^{(u)},u=1,2$, the priors are 
\begin{align*}
\begin{split}
     \psi_0^{(u)} & \sim \begin{cases} 
      N(0, 100^2)^+, & u=1 \\
      N(0, 100^2)^-, & u=2 
   \end{cases}. \\
     \end{split}
\end{align*}
Here, we use truncated priors for $\psi_0^{(1)}$ and $\psi_0^{(2)}$, as a positive dose effect and a negative squared-dose effect on the defined outcome  are substantively reasonable and  have been found in previous work \citep{schulz2021doubly}. Regarding the variance component parameters $\sigma_t^{(u)}$, a half-Cauchy (0,1) prior is used. The Bayesian hierarchical model is implemented in \texttt{RStan} \citep{stan,rstan}; 2000 posterior samples are drawn from two chains for each parameter.




\end{appendices}

\bibliographystyle{abbrvnat}
\bibliography{zsupp}
